\documentclass[utf8]{frontiersSCNS} 

\usepackage{url,lineno,microtype,subcaption}
\usepackage[onehalfspacing]{setspace}
\usepackage[colorlinks, allcolors=blue]{hyperref}

\newcommand\degr{\mbox{$^\circ$}}%
\newcommand\arcmin{\mbox{$^\prime$}}%
\newcommand\arcsec{\mbox{$^{\prime\prime}$}}%

\def\keyFont{\fontsize{8}{11}\helveticabold }
\def\firstAuthorLast{Liu {et~al.}} 
\def\Authors{Junhao Liu\,$^{1,*}$, Qizhou Zhang\,$^{2}$ and Keping Qiu\,$^{3,4}$}
\def\Address{$^{1}$East Asian Observatory, 660 N. A`oh\={o}k\={u} Place, University Park, Hilo, HI, USA \\
$^{2}$Center for Astrophysics $\vert$ Harvard \& Smithsonian, Cambridge, MA, USA\\
$^{3}$School of Astronomy and Space Science, Nanjing University, Nanjing, People's Republic of China\\
$^{4}$Key Laboratory of Modern Astronomy and Astrophysics (Nanjing University), Ministry of Education, Nanjing, People's Republic of China}
\def\corrAuthor{Junhao Liu}

\def\corrEmail{j.liu@eaobservatory.org; liujunhao42@outlook.com}

\begin{document}

\onecolumn
\firstpage{1}

\title[Magnetic fields in star formation]{Magnetic field properties in star formation: a review of their analysis methods and interpretation}


\author[\firstAuthorLast ]{\Authors} 
\address{} 
\correspondance{} 

\extraAuth{}

\maketitle

\begin{abstract}

\section{}
Linearly polarized emission from dust grains and molecular spectroscopy is an effective probe of the magnetic field topology in the interstellar medium and molecular clouds.
The longstanding Davis-Chandrasekhar-Fermi (DCF) method and the recently developed Histogram of Relative Orientations (HRO) analysis and the polarization-intensity gradient (KTH) method are widely used to assess the dynamic role of magnetic fields in star formation based on the plane-of-sky component of field orientations inferred from the observations. We review the advances and limitations of these methods and summarize their applications to observations. Numerical tests of the DCF method, including its various variants, indicate that its largest uncertainty may come from the assumption of energy equipartition, which should be further calibrated with simulations and observations. We suggest that the ordered and turbulent magnetic fields of particular observations are local properties of the considered region. An analysis of the polarization observations using DCF estimations suggests that magnetically trans-to-super-critical and averagely trans-to-super-Alfv\'{e}nic clumps/cores form in sub-critical clouds. High-mass star-forming regions may be more gravity-dominant than their low-mass counterparts due to higher column density. The observational HRO studies clearly reveal that the preferential relative orientation between the magnetic field and density structures changes from parallel to perpendicular with increasing column densities, which, in conjunction with simulations, suggests that star formation is ongoing in trans-to-sub-Alfv\'{e}nic clouds. There is a possible transition back from perpendicular to random alignment at higher column densities. Results from observational studies using the KTH method broadly agree with those of the HRO and DCF studies. 

\tiny
 \keyFont{ \section{Keywords:} Star formation, Magnetic fields, Polarization, Molecular clouds, Interstellar Medium} 
\end{abstract}

\section{Introduction}
Star formation within molecular clouds, the densest part of the interstellar medium (ISM), is regulated by the complex interplay among gravity, turbulence, magnetic fields, and other factors (e.g, protosteller feedback and feedback from previous generations of stars) at different scales. Magnetic fields interact with the other two major forces (gravity and turbulence) by providing supports against gravitational collapse \citep{1987ARA&A..25...23S} and generating anisotropic turbulence \citep{1995ApJ...438..763G}. Observational studies of magnetic fields are crucial to distinguish between strong-field star formation theories in which magnetically sub-critical clouds slowly form super-critical substructures that subsequently collapse \citep{2006ApJ...646.1043M}, and weak-field star formation theories where large-scale supersonic turbulent flows form overdense intersecting regions that dynamically collapse \citep{2004RvMP...76..125M}. 

Polarized thermal dust emission observations have been the most common way to trace the plane-of-sky (POS) component of magnetic field orientations with the assumption that the shortest axis of irregular dust grains is aligned with magnetic field lines \citep{1949PhRv...75.1605D, 2007MNRAS.378..910L, 2007JQSRT.106..225L}. The Goldreich-Kylafis (GK) effect provides an alternative way to trace the POS field orientation (with a 90$^{\circ}$ ambiguity) with molecular line polarization observations \citep{1981ApJ...243L..75G}. The recently developed Velocity Gradient Technique (VGT) proposed another way to trace the POS field orientation with line observations based on the notion that the gradient of velocity centroids \citep[VCG,][]{2017ApJ...835...41G} or thin velocity channels \citep[VChG,][]{2018ApJ...853...96L} is perpendicular to the magnetic field due to the intrinsic properties of magneto-hydrodynamic (MHD) turbulence \citep{1995ApJ...438..763G}. 

Several analysis techniques have been developed to infer the properties of magnetic fields based on their orientations: the Davis-Chandrasekhar-Fermi (DCF) method was proposed by \citet{1951PhRv...81..890D} and \citet{1953ApJ...118..113C} approximately 70 years ago and has been the most widely used method to indirectly derive the magnetic field strength with statistics of field orientations. A new tool, the polarization-intensity gradient method \citep[here after the KTH method, ][]{2012ApJ...747...79K}, was proposed about one decade ago and can also be used to assess the significance of magnetic fields based on ideal MHD equations. The Histogram of Relative Orientations (HRO) analysis \citep{2013ApJ...774..128S}, which was proposed right after the KTH method, measures the relative orientation between the magnetic field and density structures and can be used to link the observed magnetic morphology with the physics of simulations. These methods provide information on the magnetic properties in star-forming molecular clouds and allow us to investigate both qualitatively and quantitatively the dynamical role of magnetic fields in the collapse and fragmentation of dense molecular structures. 

In this chapter we review the concept and recent developments of these techniques and discuss their limitations. We also summarize the application of these methods to observations of star formation regions and discuss the role of magnetic fields at different spatial scales. In particular, we focus on the relative importance of the magnetic field as compared to gravity and turbulence at different scales of star-forming clouds. In Section \ref{sec:dcf}, we review the DCF method. In Section \ref{sec:hro}, we review the HRO analysis. In Section \ref{sec:KTH}, we review the KTH method. In Section \ref{sec:sum}, we summarize this chapter. 
\section{The DCF method}\label{sec:dcf}

In the middle 20th century, \citet{1951PhRv...81..890D} and \citet{1953ApJ...118..113C} proposed the DCF method to estimate the mean\footnote{In this paper, the mean field refers to the vector-averaged magnetic field ($B^\mathrm{m}$) and the ordered field refers to the curved large-scale ordered magnetic field ($B^\mathrm{o}$). We also use the term ``underlying field'' ($B^\mathrm{u}$) to refer to either the mean field or ordered field since many previous studies did not explicitly differ the two. The ordered field and the mean field are equivalent if the large-scale ordered field lines are straight.}  magnetic field strength ($B^\mathrm{m}$) of the interstellar medium (ISM) in the spiral arm based on the first interstellar magnetic field orientation observation made by \citet{1949Sci...109..165H}. Since then, the method has been improved and adopted by the community to estimate the field strength in star-forming regions. In this section, we present a review of the original DCF method and its modifications.

\subsection{Basic assumptions} \label{sec:assump}
\subsubsection{Energy equipartition} \label{sec:energyeq}

The original DCF method assumes an equipartition between the transverse (i.e.,  perpendicular to the underlying field $\boldsymbol{B^{\mathrm{u}}}$) turbulent magnetic and kinetic energies (i.e., the Alfv\'{e}n relation, hereafter the DCF53 equipartition assumption):
\begin{equation}\label{eq:alfven42}
\frac{1}{2} \rho  \delta v_{\mathrm{\perp}}^2= \frac{(B^{\mathrm{t}}_{\mathrm{\perp}})^2}{2 \mu_0},
\end{equation}
in SI units\footnote{The equations in SI units in this paper can be transformed to Gaussian units by replacing $\mu_0$ with $4\pi$.}, where $B^{\mathrm{t}}_{\mathrm{\perp}}$ is the transverse turbulent magnetic field, $\delta v_{\mathrm{\perp}}$ is the transverse turbulent velocity dispersion, $\mu_0$ is the permeability of vacuum, $\rho$ is the gas density. The DCF53 assumption is also adopted by the recently proposed Differential Measure Approach \citep[DMA, ][]{2022arXiv220409731L}. In the POS, the DCF53 assumption yields
\begin{equation}\label{eq:alfvenpos}
\frac{1}{2} \rho  \delta v_{\mathrm{pos\perp}}^2= \frac{(B^{\mathrm{t}}_{\mathrm{pos\perp}})^2}{2 \mu_0},
\end{equation}
where ``pos'' stands for the POS.

Alternatively, \citet{2016JPlPh..82f5301F} assumed an equipartition between the coupling-term magnetic field ($\boldsymbol{B^{\mathrm{t}}_{\mathrm{\|}}} \cdot \boldsymbol{B^{\mathrm{u}}}$, where ``$\|$'' denote the direction parallel to $\boldsymbol{B^{\mathrm{u}}}$) and the turbulence (hereafter the Fed16 equipartition assumption) when the underlying field is strong. \citet{2021AA...656A.118S} further proposed that only the transverse velocity field is responsible for $\boldsymbol{B^{\mathrm{t}}_{\mathrm{\|}}} \cdot \boldsymbol{B^{\mathrm{u}}}$ and the transverse velocity field for the POS coupling-term field can be approximated with the line-of-sight (LOS) velocity dispersion $\delta v_{\mathrm{los}}$ (see their Section 4.2). Thus they obtained
\begin{equation}\label{eq:eqcouplingpos1}
\frac{1}{2} \rho \delta v_{\mathrm{los}}^2= \frac{B^{\mathrm{t}}_{\mathrm{pos\|}} B^{\mathrm{u}}_{\mathrm{pos}}}{\mu_0},
\end{equation}
where the POS transverse velocity dispersion is neglected.  

\paragraph{Sub-Alfv\'{e}nic case}
Conventionally, a sub-Alfv\'{e}nic state means that the underlying magnetic energy is greater than the turbulent kinetic energy when comparing the magnetic field with the turbulence. It is widely accepted that the DCF53 equipartition assumption is satisfied for pure incompressible sub-Alfv\'{e}nic turbulence due to the magnetic freezing effect where the perturbed magnetic lines of force oscillate with the same velocity as the turbulent gas in the transverse direction \citep{1942Natur.150..405A}. However, the star-forming molecular clouds are highly compressible \citep{2012A&ARv..20...55H}. For compressible sub-Alfv\'{e}nic turbulence, there are still debates on whether the DCF53 or Fed16 equipartition assumptions is more accurate \citep[e.g.,][]{2021AA...647A.186S, 2021AA...656A.118S, 2022MNRAS.510.6085L, 2022ApJ...925...30L, 2022arXiv220409731L}. 

Observational studies usually adopt the local underlying field within the region of interest instead of the global underlying field at larger scales. In this case, the volume-averaged coupling-term magnetic energy is 0 by definition \citep{1995ApJ...439..779Z, 2022ApJ...925...30L}, which should not be used in analyses. Several numerical studies \citep{2016JPlPh..82f5301F, 2020MNRAS.498.1593B, 2021AA...647A.186S, 2022MNRAS.515.5267B} have suggested that the volume-averaged RMS coupling-term magnetic energy fluctuation should be studied instead of the volume-averaged coupling-term magnetic energy. With non-self-gravitating sub-Alfv\'{e}nic simulations, they found that the coupling-term field energy fluctuation is in equipartition with the turbulent kinetic energy within the whole simulation box. However, it is unclear whether the Fed16 equipartition assumption still holds in sub-regions of their simulations. Investigating the local energetics is very important because the local and global properties of MHD turbulence can be very different \citep{1999ApJ...517..700L, 2013SSRv..178..163B}. In small-scale sub-regions below the turbulent injection scale and without significant self-gravity, the local underlying magnetic field is actually part of the turbulent magnetic field at larger scales and the local turbulence is the cascaded turbulence \citep{2016JPlPh..82f5301F}. Within self-gravitating molecular clouds, the gravity is comparable to or dominates the magnetic field and turbulence at higher densities \citep{2012ARA&A..50...29C, 2013ApJ...779..185K, 2022ApJ...925...30L}, which has a strong effect on both magnetic fields and turbulence. e.g., the gravity can compress magnetic field lines and amplify the field strength; the gravitational inflows can accelerate the gas and enhance turbulent motions. As observations can only probe the magnetic field in part of the diffuse ISM or molecular clouds, it is necessary to test the validity of the Fed16 assumption in sub-regions of simulations with or without self-gravity. Moreover, \citet{2022MNRAS.510.6085L} and \citet{2022arXiv220409731L} pointed out that the physical meaning of the RMS coupling-term energy fluctuation adopted by the Fed16 equipartition assumption is still unclear, which needs to be addressed in the future. 

The traditional DCF53 equipartition assumption has been tested by many numerical works \citep[e.g.,][]{2004PhRvE..70c6408H, 2008ApJ...679..537F, 2021AA...656A.118S, 2021ApJ...919...79L, 2022MNRAS.514.1575C}. For non-self-gravitating simulations, \citet{2004PhRvE..70c6408H} found the DCF53 equipartition is violated throughout the inertial range (i.e., between the turbulence injection scale $L_{inj}$ and dissipation scale) in initially very sub-Alfv\'{e}nic ($\mathcal{M}_{A0} = 0.05-0.5$) simulations, while \citet{2008ApJ...679..537F} found an exact equipartition between turbulent magnetic and kinetic energies for initially slightly sub-Alfv\'{e}nic (initial Alfv\'{e}nic Mach number $\mathcal{M}_{A0} = 0.7$) models. Another numerical work by \citet{2021AA...656A.118S} with non-self-gravitating simulations found that the DCF53 assumption is approximately fulfilled for initially trans-Alfv\'{e}nic ($\mathcal{M}_{A0} = 0.7-2$) models, but the turbulent magnetic energy is much smaller than the turbulent kinetic energy in initially sub-Alfv\'{e}nic ($\mathcal{M}_{A0} = 0.1-0.5$) models. Similarly, as these studies adopted the whole simulation-averaged field as the mean field, it is unclear whether these relations still holds in sub-regions where the local properties are dominant. \citet{2022arXiv220409731L} suggested that the DCF53 equipartition in sub-Alfv\'{e}nic and non-self-gravitating media can be established in the regime of strong turbulence at small scales ($<L_{inj}\mathcal{M}_{A}^2$). For self-gravitating simulations, a numerical study of clustered star-forming clumps found that the DCF53 energy equipartition assumption is approximately fulfilled in trans-Alfv\'{e}nic ($\mathcal{M}_{A} = 0.7-2.5$) clumps/cores at 1-0.1 pc scales \citep{2021ApJ...919...79L}. Another numerical study by \citet{2022MNRAS.514.1575C} with self-gravitating and initially trans-Alfv\'{e}nic ($\mathcal{M}_{A0} = 1.04-1.45$) simulations found that the DCF53 equipartition approximately holds in the whole simulations box. \citet{2022MNRAS.514.1575C} further suggests that the DCF53 equipartition breaks in sub-blocks with insufficient cell numbers (e.g., $<41^3$ cells blocks). It is unclear whether the DCF53 energy equipartition assumption still holds in very sub-Alfv\'{e}nic ($\mathcal{M}_{A} \ll 0.7$) clumps/cores, although the real question may be whether there are very sub-Alfv\'{e}nic clumps/cores in gravity-accelerated gas (see discussions in Sections \ref{sec:dcfobs_BVSturb} and \ref{sec:hrosim}). 

In summary, the DCF53 equipartition assumption is valid within trans- or slightly sub-Alfv\'{e}nic self-gravitating molecular clouds, but its validity in very sub-Alfv\'{e}nic self-gravitating regions still needs more investigations. The Fed16 equipartition assumption can be used as an empirical relation when studying the global underlying and turbulent magnetic field in the diffuse ISM beyond the turbulent injection scale if the ISM is sub-Alfv\'{e}nic, but its physical interpretation, as well as its applicability in part of the ISM below the turbulent injection scale and within self-gravitating molecular clouds, are still unclear. The equipartition problem has only been investigated theoretically and numerically. We are unaware of any observational attempts to look into this problem within molecular clouds yet.  There is also a lack of observational methods to study the energy equipartition.

\paragraph{Super-Alfv\'{e}nic case}
A super-Alfv\'{e}nic state of turbulence conventionally means that the underlying magnetic energy is smaller than the turbulent kinetic energy. In super-Alfv\'{e}nic case, the magnetic field is dominated by the turbulent component. Numerical studies have shown that both the turbulent magnetic energy and the RMS coupling-term magnetic energy fluctuation are smaller than the turbulent kinetic energy in super-Alfv\'{e}nic simulations \citep{2008ApJ...679..537F, 2016JPlPh..82f5301F, 2021ApJ...919...79L}. This energy non-equipartition in super-Alfv\'{e}nic cases could lead to an overestimation of the magnetic field strength. \citet{2022arXiv220409731L} suggested that the super-Alfv\'{e}nic turbulence transfers to sub-Alfv\'{e}nic turbulence at $< L_{inj}\mathcal{M}_{A}^{-3}$ scales due to the turbulence cascade in the absence of gravity. 

\subsubsection{Isotropic turbulence}
The original DCF method assumes isotropic turbulent velocity dispersion, thus the unobserved $\delta v_{\mathrm{pos\perp}}$ in Equation \ref{eq:alfvenpos} can be replaced by the observable $\delta v_{\mathrm{los}}$, which implies 
\begin{equation}\label{eq:alfveniso}
\frac{1}{2} \rho  \delta v_{\mathrm{los}}^2= \frac{(B^{\mathrm{t}}_{\mathrm{pos\perp}})^2}{2 \mu_0}.
\end{equation}
On the other hand, the modified DCF method in \citet{2021AA...647A.186S} (hereafter ST21) requires an assumption of isotropic turbulent magnetic field, so that the $B^{\mathrm{t}}_{\mathrm{pos\|}}$ in Equation \ref{eq:eqcouplingpos1} can be replaced by $B^{\mathrm{t}}_{\mathrm{pos\perp}}$ \citep{2021AA...656A.118S}. Then there is 
\begin{equation}\label{eq:eqcouplingpos1iso}
B^{\mathrm{u}}_{\mathrm{pos}} \sim \frac{\mu_0 \rho \delta v_{\mathrm{los}}^2}{2 B^{\mathrm{t}}_{\mathrm{pos\perp}}}.
\end{equation}

However, both incompressible and compressible MHD turbulence are anisotropic in the presence of a strong magnetic field \citep{1983JPlPh..29..525S, 1984ApJ...285..109H, 1995ApJ...438..763G}. In particular, the fluctuations of the Alfv\'{e}nic and slow modes are anisotropic, while only the fast mode has isotropic fluctuations \citep{2022arXiv220409731L}. The anisotropic velocity field in low-density non-self-gravitating regions has been confirmed by observations \citep{2008ApJ...680..420H}. \citet{2022arXiv220409731L} suggested that the anisotropy of MHD turbulence is a function of $\mathcal{M}_{A}$ and relative fraction of MHD modes. The anisotropy of turbulence is also scale-dependent, which increases with decreasing scales \citep{1999ApJ...517..700L}. The recently developed DMA does not require the assumption of isotropic turbulence, but analytically considers the anisotropy of MHD turbulence in non-self-gravitating turbulent media. In particular, the DMA has written the POS transverse magnetic field structure function ($\tilde{D}_{yy}$) and the POS velocity centroid structure function ($\tilde{D}^{v}$) as functions of the turbulent field fluctuation and the velocity dispersion, respectively. Both $\tilde{D}_{yy}$ and $\tilde{D}^{v}$ are complicated functions of the mean field inclination angle\footnote{The inclination angle $\gamma$ corresponds to the angle between the 3D mean field and the LOS throughout this paper.} $\gamma$, $\mathcal{M}_{A}$, composition of turbulence modes, and distance displacements. We refer the readers to the original DMA paper \citep{2022arXiv220409731L} for the derivation of $\tilde{D}_{yy}$ and $\tilde{D}^{v}$.

With non-self-gravitating simulations, \citet{2001ApJ...561..800H} and \citet{2021AA...656A.118S} found that the full cube turbulent magnetic field is approximately isotropic within a factor of 2 with their $\mathcal{M}_{A}$=0.1-14 simulations, but they did not investigate the property of turbulence in sub-regions at smaller scales where the anisotropy is expected to be more prominent. Both works found the turbulent field is more anisotropic for models with smaller $\mathcal{M}_{A}$ values. 

Studies on the anisotropy of MHD turbulence in the high-density self-gravitating regime are rarer. Spectroscopic observations toward high-density regions did not find obvious evidences for velocity anisotropy \citep{2012MNRAS.420.1562H}. A recent numerical work by \citet{2021ApJ...919...79L} found both the turbulent magnetic field and the turbulent velocity dispersion are approximately isotropic within a factor of 2 at 1-0.01 pc scales in their trans-to-super-Alfv\'{e}nic simulations of clustered star-forming regions. They did not find obvious relation between the anisotropy and levels of initial magnetization in their simulations. This may be due to the local super-Alfv\'{e}nic turbulence at high densities and/or the complex averaging along the LOS for different local anisotropic turbulent field at various directions. Moreover, \citet{2017ApJ...836...95O} found that the velocity anisotropy due to magnetic fields disappears in high-density and sub-Alfv\'{e}nic (local $\mathcal{M}_{A}=0.54$) regions of their self-gravitating simulations. Thus, the uncertainty brought by the anisotropic MHD turbulence should be a minor contributor for the DCF method in star-forming clumps/cores where self-gravity is important.

\subsubsection{Tracing ratios between field components with field orientations}\label{sec:ratiobori}
In the POS, the local total field at position i is a vector sum of the underlying field and the local turbulent field: $\boldsymbol{B^{\mathrm{tot,i}}_{\mathrm{pos}}} = \boldsymbol{B^{\mathrm{u}}_{\mathrm{pos}}} + \boldsymbol{B^{\mathrm{t,i}}_{\mathrm{pos}}}$. Equation \ref{eq:alfveniso} can be rewritten as
\begin{equation}\label{eq:eqequitrans}
B^{\mathrm{u}}_{\mathrm{pos}} = \sqrt{\mu_0 \rho }\frac{\delta v_{\mathrm{los}}}{B^{\mathrm{t}}_{\mathrm{pos\perp}}/B^{\mathrm{u}}_{\mathrm{pos}}}.
\end{equation}
or
\begin{equation}\label{eq:eqequitranstot}
B^{\mathrm{tot}}_{\mathrm{pos}} = \sqrt{\mu_0 \rho }\frac{\delta v_{\mathrm{los}}}{B^{\mathrm{t}}_{\mathrm{pos\perp}}/B^{\mathrm{tot}}_{\mathrm{pos}}}
\end{equation}
to derive the POS underlying magnetic field strength $B^{\mathrm{u}}_{\mathrm{pos}}$ or the POS total magnetic field strength $B^{\mathrm{tot}}_{\mathrm{pos}}$. Similarly for the ST21 approach, Equation \ref{eq:eqcouplingpos1iso} can be rewritten as 
\begin{equation}\label{eq:eqequitransst21}
B^{\mathrm{u}}_{\mathrm{pos}} = \sqrt{\frac{\mu_0 \rho } {2B^{\mathrm{t}}_{\mathrm{pos\perp}}/B^{\mathrm{u}}_{\mathrm{pos}}}}\delta v_{\mathrm{los}}
\end{equation}
to derive the POS underlying magnetic field strength. Both the original DCF method and the ST21 approach have adopted the assumption that the turbulent-to-underlying field strength ratio $B^{\mathrm{t}}_{\mathrm{pos\perp}}/B^{\mathrm{u}}_{\mathrm{pos}}$ or turbulent-to-total field strength ratio $B^{\mathrm{t}}_{\mathrm{pos\perp}}/B^{\mathrm{tot}}_{\mathrm{pos}}$ in the above equations can be estimated from statistics of the POS magnetic field orientations, which is usually done by calculating the dispersion of magnetic field position angles or applying the angular dispersion function method \citep[hereafter the ADF method, ][]{2009ApJ...696..567H, 2009ApJ...706.1504H, 2016ApJ...820...38H}. On the other hand, the DMA assumes that the POS polarization angle structure function $D^{\phi}$ can be expressed as a function of the ratio between the POS transverse and total magnetic field structure functions ($\tilde{D}_{yy}/\tilde{D}_{Bpos}$), where the term $\tilde{D}_{yy}/\tilde{D}_{Bpos}$ can be expressed as a function of the turbulent-to-underlying field strength ratio. 

\paragraph{Angular dispersions}\label{sec:angles}
The underlying magnetic field reflects the intrinsic property of an unperturbed magnetic field. There are different approaches to relate the dispersion of magnetic field position angles with $B^{\mathrm{t}}_{\mathrm{pos\perp}}/B^{\mathrm{u}}_{\mathrm{pos}}$. Table \ref{tab:formulaBu} summarises these angular relations and the corresponding DCF formulas for $B^{\mathrm{u}}_{\mathrm{pos}}$. 

\begin{table}[tbh]
\tiny
\caption{Angular relations and corresponding formulas to derive $B^{\mathrm{u}}_{\mathrm{pos}}$.  \label{tab:formulaBu}}

\begin{tabular}{ccc}
\hline \noalign {\smallskip}
Relation ($B^{\mathrm{t}}_{\mathrm{pos\perp}}/B^{\mathrm{u}}_{\mathrm{pos}}\sim$)  & Formula ($ \sqrt{\mu_0 \rho }\delta v_{\mathrm{los}} \times) $ & Reference$^a$ \\
\hline \noalign {\smallskip}
$\delta\phi$ & $1/\delta\phi$ & 1,2 \\ 
$\delta (\tan \phi)$ &  $1/ \delta (\tan \phi)$ & 3,4 \\
$\tan \delta \phi$ &  $1/\tan \delta \phi$ & 5 \\
$\delta\phi$ &  $(1/\sqrt{2\delta\phi})^b$ & 6 \\  
\hline \noalign {\smallskip}
\end{tabular}

\normalsize{Notes}\\
\normalsize{$^{a}$ References: (1) \citet{1951PhRv...81..890D}; (2) \citet{1953ApJ...118..113C}; (3) \citet{2001ApJ...561..800H}; (4) \citet{2021ApJ...919...79L}; (5) \citet{2022MNRAS.510.6085L}; (6) \citet{2021AA...656A.118S}. }\\
\normalsize{$^{b}$ This formula adopts Equation \ref{eq:eqequitransst21}, while the rest of the formulas adopt Equation \ref{eq:eqequitrans}.}\\
\end{table}

All the relations listed in Table \ref{tab:formulaBu} have assumed that $B^{\mathrm{t}}_{\mathrm{pos\|}} \ll B^{\mathrm{u}}_{\mathrm{pos}}$, so the $B^{\mathrm{t}}_{\mathrm{pos\|}}$ can be neglected and the dispersion on the field angle \footnote{The field angle considers the direction of the magnetic field and is in the range of -180$\degr$ to 180$\degr$ \citep{2022MNRAS.510.6085L}, while the position angle $\phi$ only considers the orientation of the magnetic field and is in the range of -90$\degr$ to 90$\degr$. Due to the 180$\degr$ ambiguity of dust polarization observations, only the magnetic field position angle is observable.} of the magnetic field can be approximated with the dispersion of the position angle. The angular relation $B^{\mathrm{t}}_{\mathrm{pos\perp}}/B^{\mathrm{u}}_{\mathrm{pos}} \sim \delta\phi$ has adopted an additional small angle approximation (i.e., $\delta\phi \sim \tan\delta\phi \sim \delta (\tan \phi) \ll 1$ or $B^{\mathrm{t}}_{\mathrm{pos\perp}} \ll B^{\mathrm{u}}_{\mathrm{pos}}$). \citet{2001ApJ...546..980O} and \citet{2021ApJ...919...79L} found that the angle limit for the small angle approximation is $\delta\phi \lesssim 25 \degr$ in their simulations. \citet{2021ApJ...919...79L} also suggested that $\delta\phi$ can significantly underestimate $B^{\mathrm{t}}_{\mathrm{pos\perp}}/B^{\mathrm{u}}_{\mathrm{pos}}$ for large $\delta\phi$ values. For the relation $B^{\mathrm{t}}_{\mathrm{pos\perp}}/B^{\mathrm{u}}_{\mathrm{pos}}\sim \delta (\tan \phi)$, the simulations by \citet{2001ApJ...561..800H} and \citet{2021ApJ...919...79L} suggested that $\delta (\tan \phi)$ can show significant scatters due to large values of $\delta (\tan \phi)$ when $\phi \sim 90\degr$. Thus, $\delta (\tan \phi)$ is valid to trace the $B^{\mathrm{t}}_{\mathrm{pos\perp}}/B^{\mathrm{u}}_{\mathrm{pos}} $ when $\phi$ is small, and $\delta (\tan \phi)$ reduces to $\delta \phi$ in such cases. In addition, \citet{2022MNRAS.514.1575C} found that $\tan \delta \phi$ does not trace $B^{\mathrm{t}}_{\mathrm{pos\perp}}/B^{\mathrm{u}}_{\mathrm{pos}}$ well in their simulations. 
The total magnetic field is the sum of the underlying magnetic field and the turbulent magnetic field. There are also different approaches trying to relate the dispersion of magnetic field position angles with $B^{\mathrm{t}}_{\mathrm{pos\perp}}/B^{\mathrm{tot}}_{\mathrm{pos}}$. Table \ref{tab:formulaBtot} summarises these angular relations and the corresponding DCF formulas for $B^{\mathrm{tot}}_{\mathrm{pos}}$. 

\begin{table}[tbh]
\tiny
\caption{Angular relations and corresponding formulas to derive $B^{\mathrm{tot}}_{\mathrm{pos}}$.  \label{tab:formulaBtot}}

\begin{tabular}{ccc}
\hline \noalign {\smallskip}
Relation ($B^{\mathrm{t}}_{\mathrm{pos\perp}}/B^{\mathrm{tot}}_{\mathrm{pos}}\sim$)  & Formula ($ \sqrt{\mu_0 \rho }\delta v_{\mathrm{los}} \times) $ & Reference$^a$ \\
\hline \noalign {\smallskip}
... &  $(\frac{(1+3\delta (\tan \phi)^2)^{1/2}}{\delta (\tan \phi)})^b$ & 1 \\
$\tan \delta \phi$ &  $(1/\tan \delta \phi)^c$ & 2 \\
$\delta\phi$ & $1/\delta\phi$ & 3 \\ 
$\delta (\sin \phi)$ &  $1/ \delta (\sin \phi)$ & 3  \\  
$\sin(\delta \phi)$ &  $1/ \sin(\delta \phi)$ & 4  \\  
\hline \noalign {\smallskip}
\end{tabular}

\normalsize{Notes}\\
\normalsize{$^{a}$ References: (1) \citet{2001ApJ...561..800H}; (2) \citet{2008ApJ...679..537F}; (3) \citet{2021ApJ...919...79L}; (4) \citet{2022MNRAS.514.1575C}. }\\
\normalsize{$^{b}$ \citet{2001ApJ...561..800H} assumes the underlying magnetic field is along the POS. The formula derives $B^{\mathrm{tot}}_{\mathrm{3D}}$ instead of $B^{\mathrm{tot}}_{\mathrm{pos}}$. }\\
\normalsize{$^{c}$ \citet{2008ApJ...679..537F} neglected the transverse turbulent field in the total field. The formula derives $B^{\mathrm{tot}}_{\mathrm{pos\|}}$ instead of $B^{\mathrm{tot}}_{\mathrm{pos}}$. }\\
\end{table}

Similarly, all the relations listed in Table \ref{tab:formulaBtot} have approximated the dispersion on the field angle of the magnetic field with the dispersion of the position angle, which requires $B^{\mathrm{t}}_{\mathrm{pos\|}} < B^{\mathrm{u}}_{\mathrm{pos}}$. The numerical study by \citet{2021ApJ...919...79L} found that $\delta\phi$ correlates with $B^{\mathrm{t}}_{\mathrm{pos\perp}}/B^{\mathrm{u}}_{\mathrm{pos}}$ or $B^{\mathrm{t}}_{\mathrm{pos\perp}}/B^{\mathrm{tot}}_{\mathrm{pos}}$ in small angle approximation, but $\delta\phi$ estimates $B^{\mathrm{t}}_{\mathrm{pos\perp}}/B^{\mathrm{tot}}_{\mathrm{pos}}$ rather than $B^{\mathrm{t}}_{\mathrm{pos\perp}}/B^{\mathrm{u}}_{\mathrm{pos}}$ for large $\delta\phi$ values. \citet{2021ApJ...919...79L} also suggested that $\delta (\sin \phi)$ provides a better estimation of $B^{\mathrm{t}}_{\mathrm{pos\perp}}/B^{\mathrm{tot}}_{\mathrm{pos}}$ than $\delta\phi$. Due to the scatters of $\delta (\tan \phi)$, the formula $\sqrt{\mu_0 \rho }\delta v_{\mathrm{los}} \times \frac{(1+3\delta (\tan \phi)^2)^{1/2}}{\delta (\tan \phi)} $ does not correctly estimate the total magnetic field strength \citep{2001ApJ...561..800H, 2021ApJ...919...79L}. The angular relation  $B^{\mathrm{t}}_{\mathrm{pos\perp}}/B^{\mathrm{tot}}_{\mathrm{pos}}\sim \tan \delta \phi$ has not been tested by simulations yet. In addition, \citet{2022MNRAS.514.1575C} found that $\sin \delta \phi$ and $B^{\mathrm{t}}_{\mathrm{pos\perp}}/B^{\mathrm{tot}}_{\mathrm{pos}}$ are well correlated in regions where the polarization percentage is larger than 20\% of its maximum value in the synthetic polarization maps.

\paragraph{The ADF method}\label{sec:adf}
Structure functions and correlation functions have been widely used in astrophysical studies. \citet{2008ApJ...679..537F} introduced the structure function in the study of polarization position angles. Later, the ADF method was developed by \citet{2009ApJ...696..567H} to estimate the POS turbulent-to-ordered field strength ratio $B^{\mathrm{t}}_{\mathrm{pos\perp}}/B^{\mathrm{o}}_{\mathrm{pos}}$ based on the structure function of magnetic field position angles, where $B^{\mathrm{o}}$ is the ordered field. The ADF approach developed initially in \citet{2009ApJ...696..567H} (Hil09) only corrects for the large-scale ordered field structure (see Section \ref{sec:unorder} for more discussions about the ordered field). Later, the ADF technique was extended by \citet{2009ApJ...706.1504H} (Hou09) and \citet{2016ApJ...820...38H} (Hou16) to be applied to single-dish and interferometric observational data by additionally taking into account the LOS signal integration over multiple turbulent cells, the beam-smoothing effect, and the interferometer filtering effect. The Hou09 and Hou16 variants of the ADF are in the form of the auto-correlation function, which transform to the structure function in the small angle limit. Basically, the ADF of magnetic field orientations are fitted to derive $B^{\mathrm{t}}_{\mathrm{pos\perp}}/B^{\mathrm{o}}_{\mathrm{pos}}$ or $B^{\mathrm{t}}_{\mathrm{pos\perp}}/B^{\mathrm{tot}}_{\mathrm{pos}}$. The simplest ADF only accounting for the ordered field structure has the form \citep{2009ApJ...696..567H, 2009ApJ...706.1504H, 2016ApJ...820...38H}
\begin{equation} \label{eq:adforder}
1 - \langle \cos \lbrack \Delta \Phi (l)\rbrack \rangle \sim a_2' l^2 +  (\frac{B^{\mathrm{t}}_{\mathrm{pos\perp}}}{B^{\mathrm{tot}}_{\mathrm{pos}}})^2_{\mathrm{adf}},
\end{equation}
where $\Delta \Phi (l)$ is the POS angular difference of two magnetic field segments separated by a distance $l$, $a_2' l^2$ is the second-order term of the Taylor expansion of the ordered component of the correlation function, $(\frac{B^{\mathrm{t}}_{\mathrm{pos\perp}}}{B^{\mathrm{tot}}_{\mathrm{pos}}})^2_{\mathrm{adf}} = 1/(1+1/(\frac{B^{\mathrm{t}}_{\mathrm{pos\perp}}}{B^{\mathrm{o}}_{\mathrm{pos}}})^2_{\mathrm{adf}})$, and the subscript ``adf'' indicates ADF-derived parameters. The variation of the ordered field is characterised by a scale $l_d$ and it is assumed that the higher-order terms of the Taylor expansion of the ordered field do not have significant contributions to the ADF at $l<l_d$. The ADF method also assumes that the local turbulent correlation scale (see Section \ref{sec:unturb}) is smaller than $l_d$. Because $\Delta \Phi (l)$ is constrained to be within $[-90, 90]$ degrees (i.e., the field angular dispersion approximated with the position angular dispersion) and $a_2'$ is defined as positive values (i.e., positive ordered field contribution), the maximum values of the derivable integrated turbulent-to-ordered and -total strength ratio from the simplest ADF are $\sim$0.76 and $\sim$0.6 \citep{2021ApJ...919...79L}, respectively. Due to the space limit of this review, we refer readers to the original ADF papers for more complicated forms of ADFs and their detailed derivations. The validity of the ADF method is further discussed in Section \ref{sec:uncer}.

\paragraph{The DMA}
Combining Equation \ref{eq:alfven42} and the ratio between the polarization angle structure function $D^{\phi}$ and the velocity centroid structure function $\tilde{D}^{v}$, the DMA estimates the field strength as
\begin{equation}\label{eq:eqdma}
B^{\mathrm{u}}_{\mathrm{pos}} = \sqrt{\mu_0 \rho } f \frac{\tilde{D}^{v}}{D^{\phi}},
\end{equation}
where the factor $f$ is a function of $\gamma$, $\mathcal{M}_{A}$, and composition of turbulence modes. Note that the DMA assumes the velocity and magnetic field have the same scaling, therefore, $f$ does not depend on the distance displacement. \citet{2022arXiv220409731L} has listed the formula of $f$ in different physical conditions (see their Table 3). Note that their structure function $D^{\phi}(l) = \frac{1}{2}\langle 1 - \cos \lbrack 2\Delta \Phi (l)\rbrack \rangle$ is different from the one adopted by the ADF method (the left term of Equation \ref{eq:adforder}). \citet{2022arXiv220409731L} claims that $D^{\phi}(l)$ is applicable to cases of large angle fluctuations while the $1 - \langle \cos \lbrack \Delta \Phi (l)\rbrack$ adopted by the ADF method is not.

\subsection{Uncertainties in the statistics of field orientations}\label{sec:uncer}
As stated in Section \ref{sec:ratiobori}, the turbulent-to-underlying or -total magnetic field strength ratio is assumed to be traced by statistics of magnetic field position angles. Other than the uncertainty on the assumption itself, there are various effects that could introduce uncertainties in the statistics of position angles. Here we describe these effects and summarize on how they are treated in different approaches. Note that the estimation of gas density from dust emission is associated with uncertainties on the dust-to-gas ratio,
temperature, dust opacity, and source geometry \citep[e.g., ][]{1983QJRAS..24..267H, 1994A&A...291..943O}, whereas the statistics of turbulent velocity field using spectroscopic data is affected by the chemical processes and excitation conditions of particular line tracers \citep[e.g.,][]{1993prpl.conf..163V}, density fluctuations \citep[e.g.,][]{2021ApJ...910..161Y, 2022arXiv220413760Y}, and ordered velocity fields due to star-forming activities (e.g., infall, rotation, and outflows). We do not discuss the uncertainty on the gas density and turbulent velocity field in this paper as it is beyond the scope of this review. 

\subsubsection{Ordered field structure}\label{sec:unorder}

The original DCF method was derived assuming the large-scale ordered field lines are straight. For a non-linear large-scale field, the contribution from the ordered field structure can overestimate the angular dispersion that should be only attributed to turbulence. For highly ordered magnetic fields, the underlying field structure can be fitted with simple hourglass models \citep[hereafter the HG technique. e.g.,][]{2002ApJ...566..925L, 2006Sci...313..812G, 2009ApJ...707..921R, 2014ApJ...794L..18Q} or even more complex models \citep[e.g.,][]{2018ApJ...868...51M}.


\citet{2015ApJ...799...74P} (the spatial filtering technique. Hereafter the Pil15 technique) and \citet{2017ApJ...846..122P} (the unsharp masking technique. Hereafter the Pat17 technique) tried to universally derive the ordered field orientation at each position by smoothing the magnetic field position angle among neighboring positions. \citet{2017ApJ...846..122P} tested the Pat17 technique with a set of Monte Carlo simulations and found that this technique does correctly recover the true angular dispersion if the measurement error is small. By varying the smoothing size, the Pil15 and Pat17 techniques can account for the ordered structure at different scales.  

\begin{figure}[h!]
\begin{center}
\includegraphics[width=10cm]{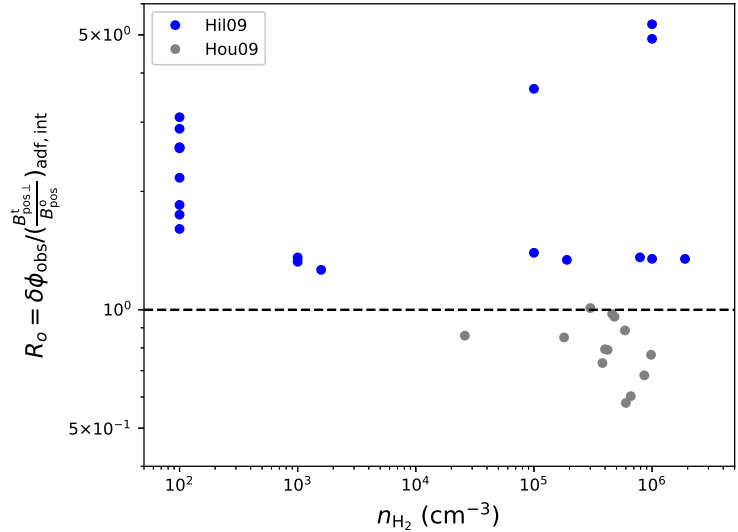}
\end{center}
\caption{Overestimation factor of angular dispersions ($R_o$) due to contributions from the ordered field structure as a function of gas number density based on the DCF catalogue from \citet{2022ApJ...925...30L} which collected the estimations in the literature. Here $\delta \phi_{\mathrm{obs}}$ is the directly estimated angular dispersion for the simplest DCF version as mentioned in Section \ref{sec:angles}. Different colors indicate different ADF variants used to derive $(\frac{B^{\mathrm{t}}_{\mathrm{pos\perp}}}{B^{\mathrm{o}}_{\mathrm{pos}}})_{\mathrm{adf,int}}$.}\label{fig:dang_btb0_n}
\end{figure}

The ADF method analytically takes into account the ordered field structure (see Section \ref{sec:adf} for detail) and has been the most widely used method to remove the contribution from the ordered field to the angular dispersion. With a set of Monte Carlo simulations, \citet{2021ApJ...919...79L} found that the ADF method works well on accounting for the ordered field. Figure \ref{fig:dang_btb0_n} shows the overestimation factor of the angular dispersion due to the POS ordered field structure, which is quantified by the ratio $R_o$ between the directly estimated angular dispersion $\delta \phi_{\mathrm{obs}}$ and the ADF-derived integrated (i.e., without corrections for the LOS signal integration. See Section \ref{sec:unlossi}) turbulent-to-ordered magnetic field strength ratio\footnote{The Hil09 variant of ADF directly estimates the $(\frac{B^{\mathrm{t}}_{\mathrm{pos\perp}}}{B^{\mathrm{o}}_{\mathrm{pos}}})_{\mathrm{adf,int}}$. For the Hou09 and Hou16 variants, the $(\frac{B^{\mathrm{t}}_{\mathrm{pos\perp}}}{B^{\mathrm{o}}_{\mathrm{pos}}})_{\mathrm{adf,int}}$ is derived by dividing the estimated $(\frac{B^{\mathrm{t}}_{\mathrm{pos\perp}}}{B^{\mathrm{o}}_{\mathrm{pos}}})_{\mathrm{adf}}$ by a factor of $\sqrt{N_{\mathrm{adf}}}$, where $N_{\mathrm{adf}}$ is the number of turbulent cells contained in the column of dust probed by the telescope beam.} $(\frac{B^{\mathrm{t}}_{\mathrm{pos\perp}}}{B^{\mathrm{o}}_{\mathrm{pos}}})_{\mathrm{adf,int}}$ from a compilation of previous DCF estimations \citep{2022ApJ...925...30L}. Figure \ref{fig:dang_btb0_n} does not show strong relations between $R_o$ and $n_{\mathrm{H_2}}$, which contradicts with the expectation that the ordered field structure is more prominent in higher-density regions where gravity is more dominant \citep[e.g.,][]{2019FrASS...6....3H}. However, most of the estimations shown in Figure \ref{fig:dang_btb0_n} were from the simplest Hil09 variant of ADF, which considers less effects and gives more uncertain results. A revisit of the data in the literature with the more refined Hou09/Hou16 approaches should give a more reliable relation between the ordered field contribution and the gas density. There is a group of data points with $R_o<1$ values derived from the Hou09 approach, which implies that the contribution from the ordered field is negative (i.e., $a_2'< 0$) and is not physical. This is because the original studies for those data points applied the Hou09 variant of ADF to interferometric data and/or fitted the ADF within an upper limit of $l$ that is too large. With Monte Carlo simulations, \citet{2019ApJ...877...43L} found that the sparse sampling of magnetic field detections can generate an artificial trend of decreasing ADF with increasing $l$ at large $l$, which can explain the unphysical $a_2'< 0$ values. The average $R_o$ for $R_o>1$ values is $2.2\pm1.1$.

A deficiency of the original ADF methods (Hil09/Hou09/Hou16) is a lack of a universal way to define the fitting upper limit of $l$ for the $a_2' l^2$ term. Therefore, users usually perform the fitting of the ADF within an arbitrary range of $l$, which can give very different results depending the adopted $l$ range. \citet{2022arXiv220409731L} suggested that the ordered field contribution to the ADF can be removed with multi-point ADFs \citep{2019ApJ...874...75C}, which has the advantage of not requiring fitting the ordered field contribution but it is at the expense of an increased noise.


It should be noted that the concept of the ``ordered'' field is vague and is not well defined. The referred entity of an local ordered field depends on the range of scales (i.e., resolution to maximum size) of the region of interest. An example is that the simple hourglass-like magnetic field in G31.41 at a lower resolution \citep{2009Sci...324.1408G} show complex poloidal-like structures at a higher resolution \citep{2019A&A...630A..54B}. It should also be noted that the non-linearly ordered field structure is not only due to non-turbulent processes such as gravity, shocks, or collisions, but can also result from larger-scale turbulence, where the curvature of the ordered field generated by pure turbulence depends on $\mathcal{M}_{A}$ \citep[e.g.,][]{2020ApJ...898...66Y}. The above mentioned techniques (except for the HG technique, where the hourglass shapes are often associated with structures under gravitational contraction) remove the contribution from non-turbulent ordered field structures as well as the contribution from the low spatial frequency turbulent field. 

\subsubsection{Correlation with turbulence}\label{sec:unturb}
It is assumed that the turbulent magnetic field is characterized by a correlation length $l_\delta$ \citep{1991ApJ...373..509M, 2009ApJ...696..567H}. The turbulent magnetic fields within a turbulent cell of length $l_\delta$ are mostly correlated with each other, while the turbulent fields separated by scales larger than $l_\delta$ are mostly independent.

\begin{figure}[h!]
\begin{center}
\includegraphics[width=10cm]{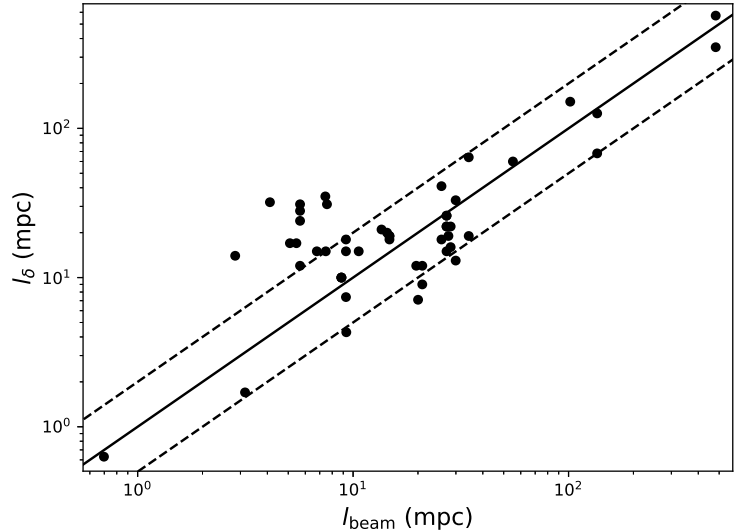}
\end{center}
\caption{The correlation length ($l_\delta$) of the turbulent magnetic field derived from the ADF method versus the beam size ($l_{\mathrm{beam}}$) of observations based on the \citet{2022ApJ...925...30L} catalogue which collected the estimations in the literature. Solid line corresponds to 1:1 relation. Dashed lines correspond to 1:2 and 2:1
relations. }\label{fig:lbeampc_ltcpc}
\end{figure}

Hou09 and Hou16 assumed a Gaussian form for the turbulent autocorrelation function and included it in the ADF analysis. Figure \ref{fig:lbeampc_ltcpc} shows the relation between $l_\delta$ derived from the ADF method and the resolution $l_{\mathrm{beam}}$ of the observations from the DCF catalogue compiled by \citet{2022ApJ...925...30L}. There is an overwhelming trend that the $l_\delta$ and $l_{\mathrm{beam}}$ are correlated with each other within a factor of 2. At $l_{\mathrm{beam}} \sim 5$ mpc, there is a group of data points with $l_\delta > 2l_{\mathrm{beam}}$, which mostly correspond to the estimations from \citet{2021ApJ...912..159P}. 

The smallest observed $l_\delta$ is $l_{\delta,\mathrm{min}}\sim$0.6 mpc estimated by \citet{2017ApJ...847...92H} in Serpens-SMM1. This scale is consistent with the lower limit of the observed ambipolar diffusion scale\footnote{Note that there is a factor of $\sqrt{3}$ difference between the correlation length of the autocorrelation funciton in \citet{2009ApJ...706.1504H} and the correlation length of a Kolmogorov-like turbulent power spectrum.} \citep[$\sim$1.8-17 mpc, ][]{2008ApJ...677.1151L, 2010ApJ...720..603H, 2011ApJ...733..109H} of ion-neutral decoupling, although recently \citet{2018ApJ...862...42T} suggested that the previous observational estimates of ambipolar diffusion scales were biased towards small values due to the imperfect method used to analyse the turbulent spectra and the true ambipolar diffusion scale may be much larger (e.g., $\sim$0.4 pc in NGC 6334). The estimated $l_{\delta,\mathrm{min}}$ is an order of magnitude smaller than the scale of the observed lower end of the Larson's law \citep[$\sim$10 mpc, ][]{2009ApJ...707L.153P, 2013ApJ...779..185K, 2022arXiv220413760Y}. 

What can we learn from the correlation between $l_\delta$ and $l_{\mathrm{beam}}$? One may intuitively think that there is an intrinsic turbulent correlation scale, which is overestimated at insufficient beam resolution. However, in this scenario, the observed angular dispersions at larger scales should be smaller than those at smaller scales due to the signal integration across more turbulent cells along the LOS (see Section \ref{sec:unlossi}), which contradicts observational results \citep{2022ApJ...925...30L}. Thus, it is reasonable to think that the magnetic field and turbulence are correlated at different scales instead of only at the smallest scale \citep{2001ApJ...561..800H}.

Alternatively, we propose that the local turbulence confined by the range of scales of the observations (from the size of the considered region, the effective depth along the LOS, or the filtering scale of interferometric observations to the beam resolution) is responsible for generating the local turbulent magnetic field at the particular scales. Note that the local ordered magnetic field partly consists of the turbulent magnetic field at larger scales and the two are not distinguishable in observations of limited scale ranges. i.e., the contribution of low spatial frequency turbulent fields can be removed by the ordered field term in the ADF analysis. In addition, coarser resolutions cannot resolve the high spatial frequency turbulence. Thus, the turbulent correlation scale derived from the ADF method actually corresponds to the local turbulent power spectra with cutoffs at the resolution and the maximum recoverable scale of particular observations. This local turbulence correlation scale in observations of molecular clouds should be much smaller than the driving scale of interstellar supersonic turbulence \citep[$\sim$100 pc, e.g.,][]{2004ApJ...615L..45H, 2008ApJ...680..362H}.

Densities may be another factor that affects the local turbulence. The property of the local turbulence relies on the gas component probed by the observations. At smaller scales, the higher densities decrease the mean free path of gas particles and thus the turbulent correlation length, which can also explain the scaling relation seen in Figure \ref{fig:lbeampc_ltcpc}. 


\subsubsection{LOS signal integration}\label{sec:unlossi}
The contribution of the turbulent field responsible for the observed polarization is the mean of the $N$ independent samples of the turbulent field along the LOS \citep{1990ApJ...362..545Z, 1991ApJ...373..509M}. The integrated turbulent field seen by observations should be $1/\sqrt{N}$ times of the intrinsic turbulent field. Note that the measured angular dispersion in polarization maps is an approximation of $B^{\mathrm{t}}_{\mathrm{pos\perp}}/B^{\mathrm{u}}_{\mathrm{pos}}$ or $B^{\mathrm{t}}_{\mathrm{pos\perp}}/B^{\mathrm{tot}}_{\mathrm{pos}}$. If both the intrinsic and integrated turbulent field dominate the underlying field, the observed and intrinsic angular dispersions should be close to the value expected for a random field and there is little difference between them. Thus, the factor $\sqrt{N}$ is only an upper limit of the underestimation factor for the angular dispersion.

\citet{2022arXiv220409731L} suggested that the underestimation of angular dispersion is dependent on the mixture of turbulence modes and the inclination of the mean field. With pure Alfv\'{e}nic simulations and simulations of equal Alfv\'{e}nic and slow modes, they found that the integrated angular dispersion decreases slower than $N^{-1/2}$ when the LOS integration depth $L$ is smaller than the turbulence injection scale $L_{inj}$, but decreases faster than $N^{-1/2}$ when $L \gtrsim L_{inj}$. They also suggested that pure Alfv\'{e}nic fluctuations decrease as $N^{-1}$ instead of $N^{-1/2}$ when the mean field is on the POS, which can quickly vanish if the integration depth is greater than the scale of the studied region. \citet{2021ApJ...919...79L} tested the LOS signal integration effect of angular dispersions with self-gravitating simulations. They found that the angular dispersion is only underestimated by a factor of $<$2 at scales $>$0.1 pc, while the angular dispersion can be significantly underestimated at scales $<$0.1 pc. However, both works only investigated the underestimation of the angular dispersion, which does not necessarily reflect the underestimation of the turbulent magnetic field. Future numerical studies should investigate the effect of LOS signal integration on the turbulent magnetic field as well as on the angular dispersion. 


The ADF method derives the POS turbulent correlation length by fitting the ADFs and uses this information to derive the number of turbulent cells $N_{\mathrm{adf}}$ along the LOS under the assumption of identical LOS and POS turbulent correlation lengths. Note that this assumption is not satisfied in anisotropic MHD turbulence. \citet{2021ApJ...919...79L} applied the ADF method to simulations and found a significant amount of observed angular dispersions corrected by $\sqrt{N_{\mathrm{adf}}}$ exceed the value expected for a random field, which they interpreted as $N_{\mathrm{adf}}$ being overestimated by the ADF method. However, as mentioned above, $\sqrt{N_{\mathrm{adf}}}$ is only an upper limit of the underestimation factor for angular dispersions. Thus their results do not necessarily mean that the ADF method inaccurately accounts for the LOS signal integration. More numerical tests of the ADF method is required to address its validity on the signal integration effect. 
\citet{2016ApJ...821...21C} proposed an alternative approach (here after the CY16 technique) to estimate the number of turbulent cells along the LOS
\begin{equation}
    N_{\mathrm{cy16}} = (\frac{\delta v_{\mathrm{los}}}{\delta V_c})^2,
\end{equation}
where $\delta V_c$ is the standard deviation of centroid velocities. The CY16 technique was developed to account for the LOS signal integration at scales larger than the injection scale in non-self-gravitation media, which does not naturally extend to small-scale and high-density media where self-gravity is important. \citet{2021ApJ...919...79L} tested the CY16 technique with self-gravitating simulations. They found that the observed angular dispersions in synthetic polarization maps corrected by $\sqrt{N_{\mathrm{cy16}}}$ agrees with the angular dispersions in simulation grids at scales $>$0.1 pc,  but the correction fails at $<$0.1 pc. The failure of the CY16 technique at $<$0.1 pc scales may be related to the complex velocity fields associated with star formation activities. Similar to the removal of ordered magnetic fields, \citet{2022arXiv220409731L} suggests that the contribution from non-turbulent velocity fields to $\delta V_c$ can be removed by multi-point structure functions. 



The non-uniform and complex density and magnetic field structures along the LOS tend to reduce the measured angular dispersion \citep[e.g., ][]{1990ApJ...362..545Z}. The distribution of the LOS grain alignment efficiency can also affect the derived angular dispersion \citep[e.g.,][]{2001ApJ...559.1005P}. The reduction factor of angular dispersions due to these effects is highly dependent on the physical conditions of individual sources and cannot be solved universally. 

\subsubsection{Observational effects}\label{sec:unobs}
The observed sky signals are limited by the angular resolution and the sampling of the particular observations. Interferometric observations are further affected by the filtering of large-scale signals. 

As discussed in Section \ref{sec:unturb}, the magnetic field is likely perturbed by the turbulence of different wavenumbers at different scales. In this case, the loss of turbulent power due to beam-smoothing can underestimate the angular dispersion of magnetic field position angles and thus overestimate the magnetic field strength \citep{2001ApJ...561..800H}. \citet{2008ApJ...679..537F} investigated the resolution effect with numerical simulations and suggested that the ratio between the derived field strength at different spatial resolutions ($R_s$) and the intrinsic field strength at an infinite resolution follows an empirical relation $(1+C/R_s^{0.5})$, where $C$ is a constant obtained by fitting the relation. It is unclear whether this empirical relation is applicable to observations or other simulations. The ADF method is another approach trying to universally solve the beam-smoothing effect, which analytically introduces a Gaussian profile to describe the telescope beam. \citet{2021ApJ...919...79L} tested the ADF method with simulations and found that this method does correctly account for the beam-smoothing effect. \citet{2021ApJ...919...79L} also suggested that the information of the turbulent magnetic field is not recoverable if the polarization source is not well resolved (i.e., size of the polarization detection area smaller than 2-4 times of the beam size). 

The minimal separation of antenna pairs in interferometric observations limits the largest spatial scale that the observations are sensitive to. This filtering effect also introduces uncertainties in the estimation of the turbulent magnetic field. The ADF method attempts to analytically solve this problem by modeling the dirty beam of interferometers with a twin Gaussian profile. With their numerical test, \citet{2021ApJ...919...79L} found the ADF method correctly accounts for this large-scale filtering effect as well. 

For observations of polarized starlight extinction or low signal-to-noise-ratio dust polarization emission, the polarization detection is sparsely sampled. \citet{2016A&A...596A..93S} has suggested that this sparse sampling effect can introduce jitter-like features in the ADFs and affect the accuracy of the ADF analysis. Although there are no universal solutions to correct for the effect of sparse sampling, the uncertainties of the ADFs due to the pure statistics can be modeled with Monte Carlo analyses \citep[e.g.,][]{2019ApJ...877...43L}. With a simple Monte Carlo test, \citet{2019ApJ...877...43L} found that the ADF of sparsely sampled magnetic field orientations is underestimated at large distance intervals and has larger scatters compared to the ADF of uniformly sampled field orientations. The sparse sampling not only affects the ADF, but can also affect the velocity structure function (VSF). \citet{2021ApJ...907L..40H} found similar jitter-like features on the VSF of sparsely sampled stars and estimated the uncertainty of the VSFs with a Monte Carlo random sampling analysis. 
 
\subsubsection{Projection effects}\label{sec:unproj}
The angular relations in Section \ref{sec:ratiobori} are originally derived in 3D spaces where the orientation of the 3D magnetic field is known. Dust polarization observations only traces the POS field orientations, thus the DCF method can only measure the POS magnetic field strength. There are different attempts to reconstruct the 3D magnetic field from the DCF estimations. 

The 3D magnetic field can be derived by including the inclination angle of the 3D field in the DCF equations \citep[e.g.,][]{2004ApJ...616L.111H, 2022arXiv220409731L}. Note that the inclination angle of the 3D field is only meaningful when the underlying field is prominent (i.e., $\mathcal{M}_{A} \lesssim 1$). \citet{2002ApJ...569..803H} proposed a technique to derive the inclination angle with the combination of the polarimetric data and ion–to–neutral line width ratios. However, \citet{2011ASPC..449..213H} later suggested that this technique cannot be widely applied due to the degeneracy between the inclination angle and the POS field strength. \citet{2019MNRAS.485.3499C} developed a technique to estimate the inclination angle from the statistical properties of the observed polarization fraction, but this technique is subject to large uncertainties (up to 30$^{\circ}$). On the other hand, \citet{2021ApJ...915...67H} suggested that the field inclination angle can be derived with the anisotropy of the intensity fluctuations in spectroscopic data. However, this approach suggested by \citet{2021ApJ...915...67H} requires sophisticated multi-parameter fittings of idealized datasets, which limits its application to observations. The applicability of their technique in self-gravitating regime is also questionable since the gravity-induced motion can significantly affect the pure velocity statistics \citep{2022MNRAS.513.2100H}. 

The 3D field strength can also be estimated by combining the POS and LOS components of the magnetic field. This can be done by combining DCF estimations and Zeeman estimations of the same material, where Zeeman observations are the only way to directly derive the LOS magnetic field strength. Recently, \citet{2018A&A...614A.100T} proposed a new method to estimate the LOS magnetic field strength based on Faraday rotation measurements, which can also be used to reconstruct the 3D magnetic field in combination with the POS magnetic field estimations \citep[e.g.,][]{2019A&A...632A..68T}.

In most cases, the information on the inclination angle and the LOS magnetic field is missing, or the measured POS and LOS magnetic fields do not correspond to the same material. One may still obtain an estimate of the 3D field from the POS field based on statistical relations.  \citet{2004ApJ...600..279C} suggested that the statistical relation between the 3D and POS underlying magnetic field is $B^{\mathrm{u}}_{\mathrm{3D}} = \frac{4}{\pi}B^{\mathrm{u}}_{\mathrm{pos}}$ for a randomly inclined field between 0 and $\pi/2$. Note that this statistical relation only applies to a collection of DCF estimations where the 3D field orientation is expected to be random, but should not be applied to individual observations. For an isotropic turbulent magnetic field, the relation between the 3D and POS turbulent field
is $B^{\mathrm{t}}_{\mathrm{3D}} = \sqrt{\frac{3}{2}}B^{\mathrm{t}}_{\mathrm{pos}}$. Therefore, the total field has the statistical relation $B^{\mathrm{tot}}_{\mathrm{3D}} = f B^{\mathrm{tot}}_{\mathrm{pos}}$, where the factor $f$ should be between $\frac{4}{\pi}\sim1.27$ and $\sqrt{\frac{3}{2}}\sim1.22$ depending on whether the underlying or turbulent field is more dominant. For an anisotropic turbulent field, the statistical relation between 3D and POS turbulent or total field should depend on the extent of the anisotropy. 


\subsection{Simulations and correlation factors}\label{sec:cor}
Due to the uncertainties in the assumptions of the DCF method (see Section \ref{sec:assump}) and on the statistics of field orientations (see Section \ref{sec:uncer}), a correction factor is required to correct for the magnetic field strength estimated from different modified DCF formulas. Several studies \citep{2001ApJ...546..980O, 2001ApJ...559.1005P, 2001ApJ...561..800H, 2008ApJ...679..537F, 2021AA...647A.186S, 2021ApJ...919...79L, 2022MNRAS.514.1575C} have numerically investigated the correction factor $Q_c$ at different physical conditions with 3D ideal compressible MHD simulations. Table \ref{tab:qcf} summarizes these numerically derived correction factors. Recently, \citet{2022arXiv220409731L} made the attempt to analytically derive the correction factor\footnote{Here we use $f$ for the analytically derived correction factor to differ with the numerically derived correction factor $Q_c$.} $f$ for their proposed DMA formula. 

\begin{table}[tbh]
\tiny
\caption{Correction factors and corresponding simulation parameters.  \label{tab:qcf}}
\begin{tabular}{cccccccc}
\hline \noalign {\smallskip}
& $\overline{Q_c}$ (range)   & Formula ($\sqrt{\mu_0 \rho }\delta v_{\mathrm{los}} \times $) & Size (pc)  & $n_{\mathrm{0}}$ or $n$ (cm$^{-3}$) & Gravity & $\mathcal{M}_{A0}$ or $\mathcal{M}_{A}$ $^f$ & Ref.$^g$ \\ 
\hline \noalign {\smallskip}

$B^{\mathrm{u}}_{\mathrm{pos}}$  & 0.5 (0.46$-$0.51)$^a$  & $1/\delta\phi_{\mathrm{obs}}$  &8   & $10^2$ & Yes & 0.7 & 1 \\ 
$B^{\mathrm{u}}_{\mathrm{pos}}$  & 0.4 (0.29$-$0.63) & $1/\delta\phi_{\mathrm{obs}}$  &1$^b$   & $< 10^5$ $^b$ & Yes & $\gtrsim$1 & 2 \\ 
$\sqrt{B^{\mathrm{u}}_{\mathrm{3D}}B^{\mathrm{tot}}_{\mathrm{3D}}}$  & (0.4$-$2.5)$^c$ &  $\frac{(1+3\delta (\tan \phi_{\mathrm{obs}})^2)^{1/4}}{\delta (\tan \phi_{\mathrm{obs}})} $  & scale-free    & ... & No & 0.8$-$14 & 3 \\ 
$B^{\mathrm{tot}}_{\mathrm{pos\|}}$  & (0.24$-$1.41)$^d$ &   $(1/\tan \delta \phi_{\mathrm{obs}})^d$ & scale-free   & ... & No & 0.7-2.0 & 4 \\
$B^{\mathrm{u}}_{\mathrm{pos}}$  & ...  & $1/\sqrt{2\delta\phi_{\mathrm{obs}}}$ & scale-free   & ... & No & 0.1-2.0 & 5 \\  
$B^{\mathrm{u}}_{\mathrm{pos}}$  & 0.28$^a$ & $1/\delta\phi_{\mathrm{obs}}$  & 1$-$0.2$^e$   & $\sim 10^4 - 10^6$ & Yes & $\sim$0.7-2.5 & 6 \\ 
$B^{\mathrm{tot}}_{\mathrm{pos}}$  & 0.62 & $1/\delta\phi_{\mathrm{obs}}$   & 1$-$0.2$^e$   & $\sim 10^4 - 10^6$ & Yes & $\sim$0.7-2.5 & 6 \\
$B^{\mathrm{tot}}_{\mathrm{pos}}$  & 0.53 & $1/\delta(\sin \phi_{\mathrm{obs}})$  & 1$-$0.2$^e$   & $\sim 10^4 - 10^6$ & Yes & $\sim$0.7-2.5 & 6 \\
$B^{\mathrm{tot}}_{\mathrm{pos}}$  & 0.21 & $(\frac{B^{\mathrm{tot}}_{\mathrm{pos}}}{B^{\mathrm{t}}_{\mathrm{pos\perp}}})_{\mathrm{adf,int}}$  & 1$-$0.2$^e$   & $\sim 10^4 - 10^6$ & Yes & $\sim$0.7-2.5 & 6 \\
$B^{\mathrm{tot}}_{\mathrm{pos}}$  & 0.39 & $(\frac{B^{\mathrm{tot}}_{\mathrm{pos}}}{B^{\mathrm{t}}_{\mathrm{pos\perp}}})_{\mathrm{adf}}$  & 1$-$0.2$^e$   & $\sim 10^4 - 10^6$ & Yes & $\sim$0.7-2.5 & 6 \\
\hline \noalign {\smallskip}
\end{tabular}

\normalsize{Notes}\\
\normalsize{$^{a}$ $\delta\phi_{\mathrm{obs}}<25 \degr$.}\\
\normalsize{$^{b}$ The simulations in \citet{2001ApJ...559.1005P} have a box size of 6.25 pc and initial $n_{\mathrm{0}} = 320$ cm$^{-3}$. They selected three 1 pc clumps for study.}\\
\normalsize{$^{c}$ After correction for the energy non-equipartition and with the assumption that the magnetic field is on the POS \citep{2001ApJ...561..800H}.}\\
\normalsize{$^{d}$ The formula in \citet{2008ApJ...679..537F} is to derive the $B^{\mathrm{tot}}_{\mathrm{pos\|}}$, but the correction factors refer to the ratio between the derived $B^{\mathrm{tot}}_{\mathrm{pos\|}}$ and the initial input $B^{\mathrm{u}}_{\mathrm{3D}}$. }\\
\normalsize{$^{e}$ The simulations in \citet{2021ApJ...919...79L} have a box size of 1-2 pc. The correction factors are derived within sub-spheres of different radii. }\\
\normalsize{$^{f}$ All the $\mathcal{M}_{A}$ were derived using the mean-field, which did not consider the ordered field structure. The ordered field contribution should be more significant for self-gravitating models.}\\
\normalsize{$^{g}$ References: (1) \citet{2001ApJ...546..980O}; (2) \citet{2001ApJ...559.1005P}; (3) \citet{2001ApJ...561..800H}; (4) \citet{2008ApJ...679..537F}; (5) \citet{2021AA...656A.118S}; (6) \citet{2021ApJ...919...79L}. For \citet{2001ApJ...559.1005P} and \citet{2021ApJ...919...79L}, the parameters reported correspond to sub-regions of the simulation at the studied time snapshot, while the parameters reported for other references are the initial parameter of the whole simulation box. }\\
\end{table}

For the most widely used formula $B^{\mathrm{u}}_{\mathrm{pos}} \sim \sqrt{\mu_0 \rho }\frac{\delta v_{\mathrm{los}}}{\delta\phi_{\mathrm{obs}}}$, \citet{2001ApJ...546..980O} made the first attempt to quantify its uncertainty and derived a correction factor of $\sim$0.5 for an initially slightly sub-Alfv\'{e}nic ($\mathcal{M}_{A0}$=0.7) model with $n_0 \sim 10^2$ cm$^{-3}$ and a box length of 8 pc. Later, \citet{2001ApJ...559.1005P} found $Q_c \sim 0.4$ for three selected $\sim$1 pc and $n < 10^5$ cm$^{-3}$ clumps in a initially slightly super-Alfv\'{e}nic model with a box length of 6.25 pc. Recently, \citet{2021ApJ...919...79L} has expanded the analysis to high-density ($\sim 10^4 - 10^6$ cm$^{-3}$) trans-Alfv\'{e}nic clumps/cores at 1-0.2 pc scales and obtained $Q_c \sim 0.28$ for several strong-field models. \citet{2021ApJ...919...79L} also found $Q_c \ll 0.28$ at $<$0.1 pc or $n > 10^6$ cm$^{-3}$ regions. Both \citet{2001ApJ...546..980O} and \citet{2021ApJ...919...79L} proposed that their correction factors are only valid when $\delta\phi_{\mathrm{obs}}<25 \degr$. Applying the same $\delta\phi_{\mathrm{obs}}<25 \degr$ criteria to the results in \citet{2001ApJ...559.1005P}, their correction factor changes to $Q_c \sim 0.31$. From the three studies, it is very clear that there is a decreasing trend of $Q_c$ with increasing density and decreasing scale in self-gravitating regions. This could be due to reduced turbulent correlation scales and more field tangling at higher densities, which leads to a more significant LOS signal averaging effect. 

\citet{2001ApJ...561..800H} proposed a formula to estimate the geometric mean of $B^{\mathrm{u}}_{\mathrm{3D}}$ and $B^{\mathrm{tot}}_{\mathrm{3D}}$ ($\sqrt{B^{\mathrm{u}}_{\mathrm{3D}}B^{\mathrm{tot}}_{\mathrm{3D}}}$, see Table \ref{tab:qcf}). However, their formula includes $\delta (\tan \phi_{\mathrm{obs}})$, which was found to have large scatters \citep{2021ApJ...919...79L}. Also, the term $\sqrt{B^{\mathrm{u}}_{\mathrm{3D}}B^{\mathrm{tot}}_{\mathrm{3D}}}$ does not have physical meaning. Thus, it is not suggested to use the geometric mean formula to estimate the magnetic field strength. 

\citet{2008ApJ...679..537F} proposed a formula $B^{\mathrm{tot}}_{\mathrm{pos\|}} \sim \sqrt{\mu_0 \rho }\frac{\delta v_{\mathrm{los}}}{\tan \delta \phi_{\mathrm{obs}}}$ and derived the correction factor for the total magnetic field strength for the first time. They compared the estimated $B^{\mathrm{tot}}_{\mathrm{pos\|}}$ with the initial input $B^{\mathrm{u}}_{\mathrm{3D}}$ in their non-self-gravitating simulations and suggested that $B^{\mathrm{tot}}_{\mathrm{pos\|}}$ is slightly larger than $B^{\mathrm{u}}_{\mathrm{3D}}$ when the magnetic field is in the POS. They also found $B^{\mathrm{tot}}_{\mathrm{pos\|}}<B^{\mathrm{u}}_{\mathrm{3D}}$ for large inclination angles. It is unclear how accurate the estimated $B^{\mathrm{tot}}_{\mathrm{pos\|}}$ values are compared to the total field strengths in their simulation. Recently, \citet{2022MNRAS.514.1575C} adopted the same formula $\sqrt{\mu_0 \rho }\frac{\delta v_{\mathrm{los}}}{\tan \delta \phi_{\mathrm{obs}}}$ but suggested that it estimates the underlying field strength than the total field strength. They also suggested that $\sqrt{\mu_0 \rho }\frac{\delta v_{\mathrm{los}}}{\sin \delta \phi_{\mathrm{obs}}}$ gives a better estimation of the total field strength. They tested the formulas with self-gravitating and initially trans-Alfv\'{e}nic ($\mathcal{M}_{A0} = 1.04-1.45$) simulations. Instead of using the area-averaged parameters ($\rho$, $\delta v_{\mathrm{los}}$, and $\delta\phi_{\mathrm{obs}}$) for the calculation, \citet{2022MNRAS.514.1575C} suggested that the average of the local pseudo-field strength calculated using the local physical parameters gives a better estimation of the true field strength. They proposed that the local gas density can be estimated with the velocity fitting method \citep{2014ApJ...794..165S} or the equilibrium method \citep{1978ApJ...220.1051E}, which, in combination with the local velocity dispersion and angular difference measurements, gives accurate field strength estimations within a factor of 2 at 1-20 pc scales and $n\sim10^{1.5} - 10^4$ cm$^{-3}$ when $\gamma>30\degr$. They also found that the correction factor decreases with increasing density and smaller scales if the analytic gas density in the simulation is adopted in the field strength estimation, which agrees with the trend found in the numerical studies of the simplest DCF formula (see discussions above in the same section). 

\citet{2021ApJ...919...79L} also investigated the correction factor for the total magnetic field strength. They found $Q_c \sim 0.62$ for $B^{\mathrm{tot}}_{\mathrm{pos}} \sim \sqrt{\mu_0 \rho }\frac{\delta v_{\mathrm{los}}}{\delta\phi_{\mathrm{obs}}}$ and $Q_c \sim 0.53$ for $B^{\mathrm{tot}}_{\mathrm{pos}} \sim \sqrt{\mu_0 \rho }\frac{\delta v_{\mathrm{los}}}{\delta (\sin \phi_{\mathrm{obs}})}$. Although $\sqrt{\mu_0 \rho }\frac{\delta v_{\mathrm{los}}}{\delta\phi_{\mathrm{obs}}}$ is often used to derive the underlying field strength, it is more likely correlated with the total field strength when the angular dispersion is large. Note that the correction factors in \citet{2021ApJ...919...79L} only apply to scales of clumps and cores when densities are greater than 10$^4$ cm$^{-3}$ and are not applicable at larger scales (e.g., ISM or clouds.). Also note that the simulations in \citet{2021ApJ...919...79L} have physical conditions similar to clustered star-forming regions. It is possible that the correction factor for isolated low-mass star formation regions could be larger than those reported by \citet{2021ApJ...919...79L} at the same scales due to lower densities. \citet{2021ApJ...919...79L} did not find significant difference among the correction factors for different inclination angles. 

\citet{2021AA...647A.186S} and \citet{2021AA...656A.118S} proposed a new formula $B^{\mathrm{u}}_{\mathrm{pos}} = \sqrt{\frac{\mu_0 \rho } {2\delta\phi_{\mathrm{obs}}}} \delta v_{\mathrm{los}}$ and tested their formula with scale-free non-self-gravitating models. They claimed that their formula does not need a correction factor. However, several assumptions in the derivation of the ST21 formula are approximations (see Section \ref{sec:assump}) and there are also uncertainties in statistics of magnetic field position angles (see Section \ref{sec:uncer}). Therefore correction factors should be still required for this method to compensate these uncertainties. More tests are needed to understand the validity of the ST21 formula under different physical conditions (e.g., at high-density self-gravitating medium). 

Other than in situations when the ADF method is improperly applied to observations (e.g., obtaining negative ordered field contribution or only adopting $a_2' l^2$ when $l>l_d$), the uncertainty of the ADF method may mainly come from the maximum derivable values of the integrated turbulent-to-ordered or -total field strength ratio (see Section \ref{sec:adf}), which underestimates the turbulent-to-ordered or -total field strength ratio and overestimates the field strength. \citet{2021ApJ...919...79L} has estimated the correction factor for the total field strength derived from the ADF method in trans-Alfv\'{e}nic clumps/cores. They failed to derive the strength of the non-linearly ordered field in their simulations, thus they did not compare the ADF-derived ordered field strength with simulation values.

A recent and important modification of the DCF method is the DMA \citep{2022arXiv220409731L}, which theoretically derives the formula $B^{\mathrm{u}}_{\mathrm{pos}} = \sqrt{\mu_0 \rho } f \frac{\tilde{D}^{v}}{D^{\phi}}$ in the non-self-gravitating regime. The analytical correction factor $f$ is a function of $\gamma$, $\mathcal{M}_{A}$, and fraction of turbulence modes. \citet{2022arXiv220409731L} has listed the asymptotic forms of $f$ and the DMA formula in typical conditions of the ISM and molecular clouds (see their Table 3). They tested the DMA formula with a set of non-self-gravitating simulations and found the analytically and numerically derived correction factors agree well with each other in typical interstellar conditions. They suggested a pronounced dependence of $f$ on the mean field inclination angle and fraction of turbulence modes in molecular clouds. However, both parameters are difficult to obtain observationally, which makes it difficult to apply the DMA to observational data. A further extension of the DMA by including self-gravity is essential to increase its accuracy in the determination of magnetic field strengths in self-gravitating molecular clouds. 

\subsection{Observational DCF estimations in star-forming regions}\label{sec:dcfobs}

The original DCF method and its modified forms have been widely applied to observations of magnetic fields in star-forming regions to estimate the magnetic field strength. Statistical studies of DCF estimations are of significant value to extend our understanding of the role of magnetic fields in star formation \citep[e.g.,][]{2019FrASS...6...15P, 2021ApJ...917...35M}. Recently, \citet{2022ApJ...925...30L} compiled all the previous DCF estimations published between 2000 and June 2021 from polarized dust emission observations within star-forming molecular clouds. Similarly, \citet{2022arXiv220311179P} made a compilation of all types of DCF measurements published between 2001 and May 2021. Here we briefly summarise the previous observational DCF estimations. 
 
\subsubsection{Comparing magnetic field with gravity}

\paragraph{$B-n$ relation}
During the contraction of molecular clouds, the gravity can compress and amplify the magnetic field. The power-law index of the $B-n$ relation ($B \propto n^{j}$) characterise the dynamical importance of magnetic field during the cloud collapse. In the case of extremely weak magnetic fields where gas collapses isotropically due to magnetic freezing, there is relation $B \propto n^{2/3}$ for a self-gravitating cloud core during its temporal evolution \citep{1966MNRAS.133..265M}. In such case, the radial component of the magnetic field also has the 2/3 scaling dependence on the gas density at any time snapshots, whereas the tangential component does not follow this scaling relation. For stronger fields, the density increases faster than the magnetic field due to ambipolar diffusion at higher densities, which results in shallower power-law slopes \citep[e.g., $j\lesssim0.5$,  ][]{1999ASIC..540..305M}.


However, the temporal $B-n$ relation of a single cloud is not obtainable observationally due to the long evolutionary time scale. Studies of the spatial $B-n$ relation for a single cloud \citep[e.g.,][]{2015Natur.520..518L} are also rare. Instead, observational $B-n$ studies usually focus on the spatial $B-n$ relation for an ensemble of star-forming regions at different evolution stages and different scales. \citet{2010ApJ...725..466C} made a pioneering study of the spatial $B-n$ relation based on the Bayesian analysis of a collection of Zeeman observations. They found that the magnetic field does not scale with density at $n_{\mathrm{H_2}} < 300$ cm$^{-3}$, but scales with density as $B \propto n^{0.65}$ at $n_{\mathrm{H_2}} > 300$ cm$^{-3}$. \citet{2021ApJ...917...35M} compiled the DCF estimation in 17 dense cores and reported $B \propto n^{0.66}$. With compilations of DCF estimations, \citet{2022ApJ...925...30L} and \citet{2022arXiv220311179P} found a similar trend to the Zeeman results in that the magnetic field does not scale with density at lower densities, but scales with density at higher densities. Due to the large scatters and the uncertainty in correction factors, they did not report the critical density and magnetic field strength for the transition. \citet{2022ApJ...925...30L} reported $B \propto n^{0.57}$ with a simple power-law fit for the high-density regime.

Despite the progress in the observational $B-n$ studies, concerns have been raised on whether the $B-n$ relation from a collection of different sources can be compared with model predictions for individual sources \citep{2021Galax...9...41L}. For the Zeeman observations, \citet{2015MNRAS.451.4384T} and \citet{2020ApJ...890..153J} found that adopting different observational uncertainties of $n$ other than the factor of 2 uncertainty adopted by \citet{2010ApJ...725..466C} can affect the fitted slope of the $B-n$ relation, which questions the validity of the Bayesian analysis in \citet{2010ApJ...725..466C}. \citet{2015MNRAS.451.4384T} further found that the samples collected in \citet{2010ApJ...725..466C} are preferentially non-spherical, which is inconsistent with the $B \propto n^{2/3}$ scaling. The DCF-derived $B - n$ relation is also very uncertain due to the scatters on the DCF estimations and the intrinsic $B \propto n^{0.5}$ dependence of the DCF method. We do not aim to present a detailed review of the $B-n$ relation, thus we refer readers to \citet{2015MNRAS.451.4384T}, \citet{2019FrASS...6...66C}, \citet{2019FrASS...6....5H}, \citet{2021Galax...9...41L}, and \citet{2022arXiv220311179P} for additional detailed discussions. 


\paragraph{Mass-to-flux-ratio to critical value} \label{sec:dcflambda}
The relative importance between the magnetic field and the gravity of individual sources is usually parameterized by the magnetic critical parameter $\lambda$ (i.e., mass-to-flux-ratio in units of the critical value). The critical value of the mass-to-flux-ratio is model-dependent \citep[e.g., ][]{1966MNRAS.132..359S, 1976ApJ...210..326M, 1978PASJ...30..671N}. The magnetic critical parameter for an isothermal disk-like structure is given by \citep{1978PASJ...30..671N, 2004ApJ...600..279C}
\begin{equation}\label{eq:lambdac04}
\lambda = \mu_{\mathrm{H_2}} m_{\mathrm{H}} \sqrt{\mu_0\pi G} \frac{N_{\mathrm{H_2}}}{B} \sim 7.6 \times 10^{-21} \frac{N_{\mathrm{H_2}}/(\mathrm{cm}^{-2})}{B/(\mu G)},
\end{equation}
where $G$ is the gravitational constant, $\mu_{\mathrm{H_2}}$ is the mean molecular weight per hydrogen molecule, and $m_{\mathrm{H}}$ is the atomic mass of hydrogen. The magnetic critical parameter for a spherical structure with a density profile of $n \propto r^{-i}$ is given by \citep{2022ApJ...925...30L}
\begin{equation} \label{eq:lambliu22}
\lambda = \mu_{\mathrm{H_2}} m_{\mathrm{H}} \sqrt{1.5\mu_0\pi G/k_i} \frac{N_{\mathrm{H_2}}}{B},
\end{equation}
where  $k_i = (5-2i)/(3-i)$. Equation \ref{eq:lambliu22} reduces to $\lambda \sim 8.7 \times 10^{-21} \frac{N_{\mathrm{H_2}}/(\mathrm{cm}^{-2})}{B/(\mu G)}$ when $i\sim$1.83 \citep{2022ApJ...925...30L}. $\lambda > 1$ indicates the gravity dominates the magnetic field (i.e., magnetically super-critical), and vice versa. Alternatively, the importance between the magnetic field and the gravity can also be compared with the magnetic virial parameter $\alpha_{\mathrm{B}} = 1/\lambda $ or with their energy ratios. The magnetic critical parameter or the magnetic virial parameter can also be expressed as a function of the number density $n_{\mathrm{H_2}}$ and radius $r$ through the relation $N_{\mathrm{H_2}} = 4n_{\mathrm{H_2}}r/3$. 


\begin{figure}[h!]
\begin{center}
\includegraphics[width=10cm]{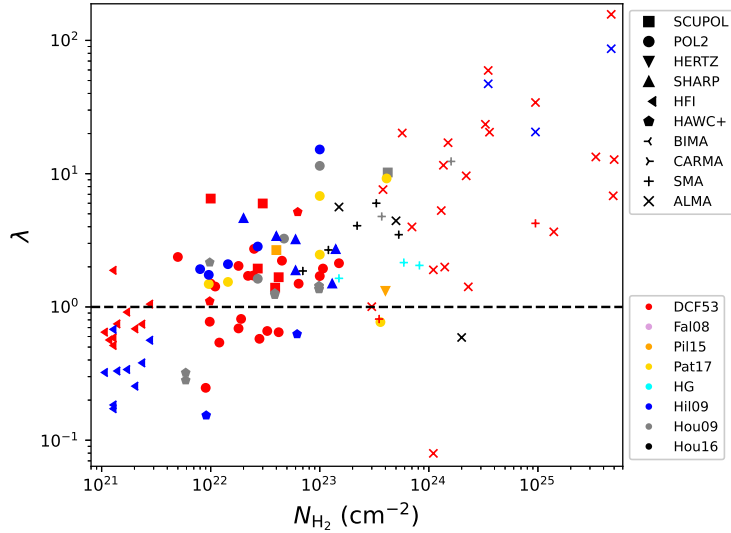}
\end{center}
\caption{The relation between the magnetic critical parameter (Equation \ref{eq:lambliu22}) and column density from the previous DCF estimations \citep{2022ApJ...925...30L}. Different symbols represent different instruments (see \citet{2022ApJ...925...30L} and references therein). Different colors represent different methods. To avoid an assumption on whether the underlying or turbulent field is more dominant, the magnetic critical parameter is estimated with the 3D total magnetic field strength adopting the statistical relation $B^{\mathrm{tot}}_{\mathrm{3D}} = 1.25 B^{\mathrm{tot}}_{\mathrm{pos}}$ (i.e., $(\sqrt{\frac{3}{2}} + 4/\pi)/2 \sim 1.25$, see Section \ref{sec:unproj}) instead of $B^{\mathrm{tot}}_{\mathrm{3D}} = \sqrt{\frac{3}{2}}B^{\mathrm{tot}}_{\mathrm{pos}}$ adopted in \citet{2022ApJ...925...30L}. }\label{fig:lamb_Ncol}
\end{figure}



Statistical DCF studies have suggested that while molecular clouds are magnetically sub-critical \citep{2016AA...586A.138P}, the molecular dense cores within clouds are super-critical \citep{2021ApJ...917...35M}. The recent more complete DCF compilation by \citet{2022ApJ...925...30L} found a clear trend of increasing $\lambda$ with increasing $N_{\mathrm{H_2}}$ (see Figure \ref{fig:lamb_Ncol}), where the average state transits from sub-critical to super-critical at $\sim3.4 \times 10^{21}$ cm$^{-2}$. This trend appears to agree with the prediction of magnetic field controlled star formation theories \citep{2006ApJ...646.1043M}, where magnetically sub-critical molecular clouds gradually form trans-to-super-critical substructures that collapse. The collapse is more dynamical in higher density regions. The dissipation of magnetic flux at higher densities may be due to ambipolar diffusion \citep{1999ASIC..540..305M} or magnetic reconnection \citep{1999ApJ...517..700L}. Mass accumulation along magnetic field lines can also increase the mass-to-flux ratio at higher densities. Despite the general trend seen in Figure \ref{fig:lamb_Ncol}, the samples collected by \citet{2022ApJ...925...30L} are mostly from different regions. Future multi-scale studies of the same region would be of great significance. High-mass star-forming regions tend to have higher column densities than low-mass star-forming regions at the same scales \citep{2014prpl.conf..149T}, thus high-mass star formation may be more magnetically super-critical than low-mass star formation within molecular clouds. There is only one DCF estimation of high-mass star-forming clouds at 10 pc scales so far \citep[Orion, ][]{2016AA...586A.138P}. More DCF estimations would be helpful for us to better understand the dynamical states of massive star formation at cloud scales. With better calibrated modified DCF methods (e.g., the extension of the DMA to the self-gravitating regime) and observational constraints for the physical parameters required for the DCF estimations, future magnetic criticality studies in molecular clouds could shed light on constraining the critical density in specific clouds and on comparing the mass-to-flux ratio of sources at different evolutionary stages.

\subsubsection{Comparing magnetic field with turbulence}\label{sec:dcfobs_BVSturb}

The relative importance between the underlying magnetic field and the turbulence of individual sources is usually parameterized by the Alfv\'{e}nic Mach number
\begin{equation}\label{eq:alfven}
\mathcal{M}_{A} = \frac{\delta v_{\mathrm{3D}}\sqrt{\mu_0 \rho }}{B^{\mathrm{u}}_{3D}}.
\end{equation}
If there is an equipartition between the turbulent magnetic and kinetic energies, Equation \ref{eq:alfven} reduces to $\mathcal{M}_{A} = B^{\mathrm{t}}_{\mathrm{3D}}/B^{\mathrm{u}}_{3D}$. The ratio $B^{\mathrm{t}}_{\mathrm{3D}}/B^{\mathrm{u}}_{3D}$ can be derived with the statistics of the observed polarization angles if the mean field inclination angle and the anisotropy of the turbulent field are known \citep{2022arXiv220409731L}. With the relations $B^{\mathrm{u}}_{\mathrm{3D}} = f_u B^{\mathrm{u}}_{\mathrm{pos}}$ and $B^{\mathrm{t}}_{\mathrm{3D}} = f_t B^{\mathrm{t}}_{\mathrm{pos\perp}}$, there is $\mathcal{M}_{A} \sim (f_t/f_u) B^{\mathrm{t}}_{\mathrm{pos\perp}}/B^{\mathrm{u}}_{\mathrm{pos}}$, where $f_u$ and $f_t$ are factors for the 3D to POS conversion. In small angle approximation, we obtain $\mathcal{M}_{A} \sim (f_t/f_u) \delta\phi_{\mathrm{obs}}$. The term $B^{\mathrm{t}}_{\mathrm{pos\perp}}/B^{\mathrm{u}}_{\mathrm{pos}}$ can also be derived from the ADF method (see Section \ref{sec:ratiobori}).
The relation between $\mathcal{M}_{A}$ and $B^{\mathrm{t}}_{\mathrm{pos\perp}}/B^{\mathrm{u}}_{\mathrm{pos}}$ or $\delta\phi_{\mathrm{obs}}$ can be regarded as an extension of the DCF formula, thus the correction factors for the DCF formula should also be applied to $\mathcal{M}_{A}$ under the same equipartition assumption. i.e., we have $\mathcal{M}_{A} \sim (f_t/f_u/Q_c) (B^{\mathrm{t}}_{\mathrm{pos\perp}}/B^{\mathrm{u}}_{\mathrm{pos}})_{\mathrm{adf}}$ or $\mathcal{M}_{A} \sim (f_t/f_u/Q_c) \delta\phi_{\mathrm{obs}}$. Adopting an additional correction factor $f_o$ to account for the ordered field contribution to the angular dispersion, the relation between $\mathcal{M}_{A}$ and $\delta\phi_{\mathrm{obs}}$ becomes $\mathcal{M}_{A} \sim (f_t/f_u/Q_c/f_o) \delta\phi_{\mathrm{obs}}$. 



\begin{figure}[h!]
\begin{center}
\includegraphics[width=10cm]{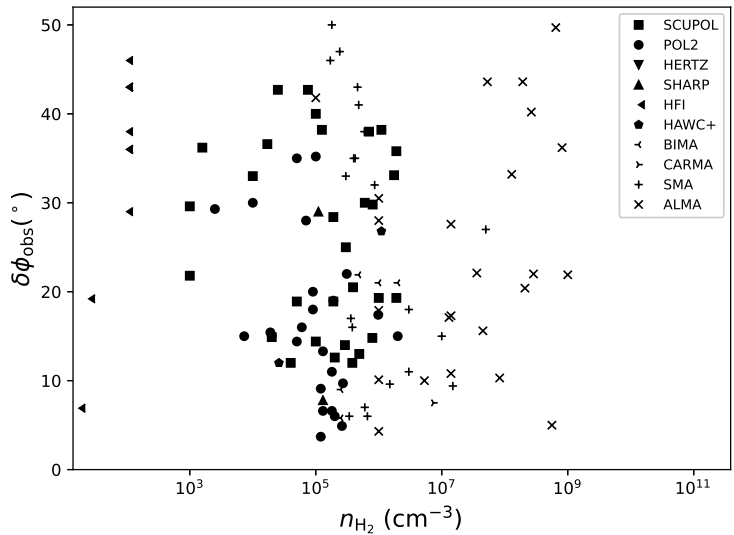}
\end{center}
\caption{The relation between the directly estimated angular dispersion and the number density from the previous DCF estimations \citep{2022ApJ...925...30L}, reproduced with permission. Different symbols represent different instruments. Some data points with $\delta\phi_{\mathrm{obs}} > 52 \degr$ are unphysical and are not plotted. }\label{fig:angdcf_dens}
\end{figure}

\citet{2022ApJ...925...30L} has compiled the observed angular dispersion from previous DCF studies (see Figure \ref{fig:angdcf_dens}). They suggested that the observed angular dispersion does not provide too much information on the Alfv\'{e}nic states of molecular clouds due to the maximum angle limit of a random field ($\delta\phi_{\mathrm{obs}} < 52 \degr$) and the lack of an appropriate DCF correction factor at cloud scales. Without knowledge of the inclination angle and turbulence anisotropy for most observations in the compilation, they adopted the statistical correction factor $f_u\sim4/\pi$ for a randomly distributed 3D mean field orientation \citep{2004ApJ...600..279C}, $f_t\sim\sqrt{3}$ for approximate isotropic turbulent fields in self-gravitating regions \citep{2021ApJ...919...79L}, and $f_o\sim2.5$ based on the numerical work by \citet{2021ApJ...919...79L}. Additionally adopting the $Q_c$ for different modified DCF approaches reported in \citet{2021ApJ...919...79L}, \citet{2022ApJ...925...30L} found that the average $\mathcal{M}_{A}$ for the substructures in molecular clouds is $\sim$0.9, which indicates the average state is approximately trans-Alfv\'{e}nic. They also suggested that both sub- and super-Alfv\'{e}nic states exist for the cloud substructures and did not find strong relations between $\mathcal{M}_{A}$ and $n$. \citet{2022arXiv220311179P} did a similar study of $\mathcal{M}_{A}$ with their compilation of observed angular dispersions. Unlike \citet{2022ApJ...925...30L}, \citet{2022arXiv220311179P} corrected the estimated $\mathcal{M}_{A}$ with the ratio between the DCF-derived POS field strength and Zeeman-derived LOS field strength at similar densities assuming that the Zeeman estimations are accurate references. They found that the turbulence is approximately trans-Alfv\'{e}nic on average and $\mathcal{M}_{A}$ has no clear dependence on $n$, which agree with \citet{2022ApJ...925...30L}. Note that both \citet{2022ApJ...925...30L} and \citet{2022arXiv220311179P} are statistical studies, where the adopted statistical relations may not apply to individual sources. 

As the ADF method removes the contribution from the ordered field, the turbulent-to-ordered field strength ratio derived by the ADF method should be more suitable for the study of the Alfv\'{e}nic state than the directly estimated angular dispersions. However, the applicability of the ADF method to determine the $\mathcal{M}_{A}$ value is limited by the maximum derivable turbulent-to-ordered field strength ratio (see Section \ref{sec:adf}), its uncertainty on the LOS signal integration (see Section \ref{sec:unlossi}), and the lack of appropriate numerically-derived correction factors for these uncertainties (see Section \ref{sec:cor}).

If an alternative assumption of the Fed16 equipartition (see Section \ref{sec:energyeq}) is adopted, the $\mathcal{M}_{A}$ should be correlated with $\sqrt{B^{\mathrm{t}}_{\mathrm{pos\perp}}/B^{\mathrm{u}}_{\mathrm{pos}}}$ or $\sqrt{\delta\phi_{\mathrm{obs}}}$ instead of $B^{\mathrm{t}}_{\mathrm{pos\perp}}/B^{\mathrm{u}}_{\mathrm{pos}}$ or $\delta\phi_{\mathrm{obs}}$ \citep{2021AA...656A.118S}. However, the applicability of these alternative relations is limited by the lack of correction factors at different physical conditions. 

In summary, the average state of star-forming substructures within molecular clouds may be approximately trans-Alfv\'{e}nic, but the observed angular dispersions do not yield clues on the Alfv\'{e}nic state of molecular clouds themselves. Note that the equipartition assumption (either the DCF53 or the Fed16 assumption), which should be independently confirmed, is a prerequisite for using the angular dispersion to determine the Alfv\'{e}nic Mach number. If the equipartition assumption is not satisfied for some of the sources, the average state should be more super-Alfv\'{e}nic. 

\subsubsection{Equilibrium state}
The equilibrium state of a dense structure is usually parameterized by the virial parameter. Neglecting the surface kinetic pressure and the thermal pressure, the total virial parameter for a spherical structure considering the support from both the magnetic field and the turbulence is estimated as the ratio of the total virial mass and the gas mass
\begin{equation}\label{eq:virialtotal}
\alpha_{\mathrm{turb+B}} = \frac{M_{\mathrm{turb+B}}}{M}.
\end{equation}
The total virial mass is given by \citep{2020ApJ...895..142L}
\begin{equation}
M_{\mathrm{turb+B}} = \sqrt{M^2_{\mathrm{B}} + (\frac{M_{\mathrm{turb}}}{2})^2} + \frac{M_{\mathrm{turb}}}{2},
\end{equation}
where the magnetic virial mass is estimated with
\begin{equation}
M_{\mathrm{B}} = \frac{\pi r^2 B}{\sqrt{1.5\mu_0\pi G/k_i}},
\end{equation}
and the turbulent virial mass is  estimated with
\begin{equation}
M_{\mathrm{turb}} = \frac{3k_i\delta v_{\mathrm{los}}^2r}{G}.
\end{equation}
Alternatively, the equilibrium state can also be studied by comparing $2E_{\mathrm{turb}} + E_{\mathrm{B}}$ with $E_{\mathrm{G}}$, where $E_{\mathrm{turb}}$ is the turbulent kinetic energy, $E_{\mathrm{B}}$ is the magnetic energy, and $E_{\mathrm{G}}$ is the gravitational energy. 

\begin{figure}[h!]
\begin{center}
\includegraphics[width=10cm]{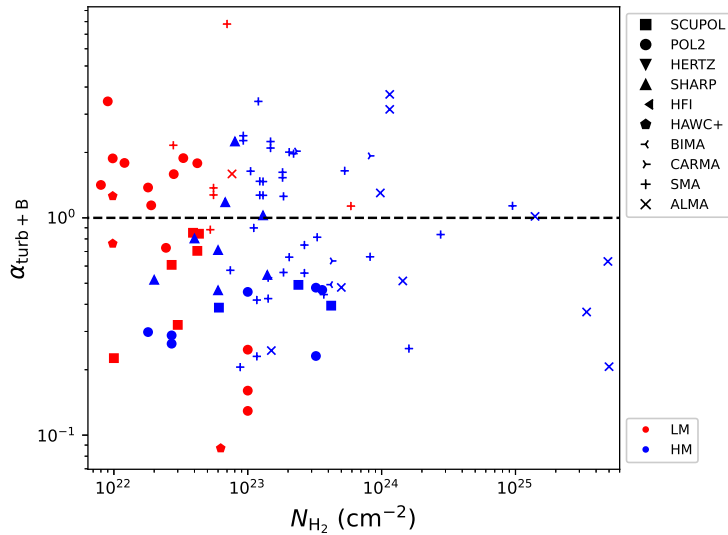}
\end{center}
\caption{The relation between the total virial parameter (Equation \ref{eq:virialtotal}) and column density for the DCF sample catalogued by \citet{2022ApJ...925...30L}. Different symbols represent different instruments. Red and blue colors indicate low-mass (LM) and high-mass (HM) star-forming regions, respectively. }\label{fig:alpha_Ncol}
\end{figure}

Figure \ref{fig:alpha_Ncol} shows the total virial parameter as a function of column density for the dense substructures within molecular clouds based on the DCF compilation by \citet{2022ApJ...925...30L}. Due to the lack of mass estimations, we do not show the $\alpha_{\mathrm{turb+B}}$ for the molecular clouds observed by the Planck. Since the magnetic field can solely support the clouds (see Section \ref{sec:dcflambda}), these clouds should have $\alpha_{\mathrm{turb+B}}>1$ (i.e., super-virial). Low-mass star-forming regions with $M \leq M_{\mathrm{crit}} = 870 M_{\odot} (r/pc)^{1.33}$ and high-mass star-forming regions with $M > M_{\mathrm{crit}}$ \citep{2010ApJ...716..433K} are indicated with different colors. In Figure \ref{fig:alpha_Ncol}, the high-mass regions with highest column densities (e.g., $N_{\mathrm{H_2}}> 10^{24}$ cm$^{-2}$) tend to be trans- or sub-virial, but both super- and sub-virial states exist at lower column densities. The median $\alpha_{\mathrm{turb+B}}$ in low-mass and high-mass regions are $\sim$1.1 and $\sim$0.66, respectively, suggesting that the gravity may be more dominant in high-mass star formation. It may also indicate that high-mass star formation within molecular clouds tends to be more likely in non-equilibrium. It is possible that the magnetic field strength is overestimated for some sources due to the energy non-equipartition, which suggests a even more dynamical massive star formation. In summary, it appears that star-forming regions with higher column densities may have smaller total virial parameters due to more significant roles of gravity, but this trend is highly uncertain due to the large scatters. 
 
\section{The HRO analysis}\label{sec:hro}

\subsection{Basics}
\citet{2013ApJ...774..128S} developed the HRO analysis to characterize the relative orientation of the magnetic field with respect to the density structures, which can be used to establish a link between observational results and the physics in simulations. In 3D, the HRO is the relation between the 3D magnetic field orientation and the number density gradient $\nabla n$. In the POS, the HRO is the relation between the POS magnetic field orientation and the column density gradient $\nabla N$. The calculation of the density gradient is by applying Gaussian derivative kernels to the density structure \citep{2013ApJ...774..128S}. The density gradient at different scales can be studied by varying the size of the derivative kernels. The orientation of the iso-density structure is perpendicular to the direction of the density gradient. 
For observational data in the POS, the angle $\phi_{\mathrm{B-N}}$ between the orientation of the iso-column density structure and the POS magnetic field orientation\footnote{Note that there is a 90$\degr$ difference between the $\phi_{\mathrm{B-N}}$ defined in \citet{2013ApJ...774..128S} and in the subsequent HRO papers. In \citet{2013ApJ...774..128S}, $\phi_{\mathrm{B-N}}$ is defined as the angle between the column density gradient and the POS magnetic field orientation.} is given by
\begin{equation}
\phi_{\mathrm{B-N}} = \arctan (\vert \nabla N\times \boldsymbol{E} \vert, \nabla N\cdot\boldsymbol{E}),
\end{equation}
where $\boldsymbol{E}$ is the POS electric field pseudo-vector  \citep{2016AA...586A.138P}. The angle $\phi_{\mathrm{B-N}}$ is within $\pm 90\degr$. When $\phi_{\mathrm{B-N}} = 0 \degr$, the magnetic field is parallel to the orientation of the column density structure and is perpendicular to the column density gradient. When $\vert\phi_{\mathrm{B-N}}\vert = 90 \degr$, the magnetic field is perpendicular to the orientation of the column density structure and is parallel to the column density gradient. The preferential relative orientation between the magnetic field and the density structure is characterized by the histogram shape parameter \citep{2016AA...586A.138P, 2017AA...603A..64S}
\begin{equation}
\xi = \frac{A_0 - A_{90}}{A_0 + A_{90}},
\end{equation}
where $A_0$ is the percentage of pixels with $\vert\phi_{\mathrm{B-N}}\vert < 22.5 \degr$ and $A_{90}$ is the percentage of pixels with $67.5 \degr < \vert\phi_{\mathrm{B-N}}\vert < 90 \degr$ in a polarization map. Positive $\xi$ values indicate the column density structure is more likely aligned with the magnetic field, and vice versa. The $\xi$ in 3D can be derived similarly. 

Using the parameter $\xi$ to characterize the relative orientation has some drawbacks. For instance, the derivation of the parameter $\xi$ completely ignores angles within $22.5 \degr < \vert\phi_{\mathrm{B-N}}\vert < 67.5 \degr$. It also suffers from intrinsic deficiencies of binning of angles. To overcome these shortcomings, \citet{2018MNRAS.474.1018J} improved the HRO analysis with the projected Rayleigh statistic (PRS), which uses the PRS parameter $Z_x$ instead of $\xi$ to characterize the preferential relative orientation. For a set of $n'$ angles of relative orientation $\{\phi_{B-N,i}\}$, $Z_x$ is estimated with
\begin{equation} \label{eq:prs}
Z_x = \frac{\Sigma^{n'}_i \cos 2\phi_{B-N,i}}{\sqrt{n'/2}}.
\end{equation}
$Z_x>0$ indicates a preferential parallel alignment between the magnetic field and column density structure, and vice versa. \citet{2018MNRAS.474.1018J} suggested that the parameter $Z_x$ is more statistically powerful than the parameter $\xi$, especially when the sample size is small or when the angles $\{\phi_{B-N,i}\}$ are more uniformly distributed. Equation \ref{eq:prs} cannot be directly applied to 3D data. The formula for the 3D PRS parameter still needs more theoretical investigations \citep{2021MNRAS.503.5425B}. 

The VGT group \citep[e.g.,][]{2017ApJ...835...41G, 2018ApJ...853...96L} introduced an alignment measure ($AM$) parameter to study the relative orientation between magnetic field orientations and velocity gradients, where the $AM$ can also be used to study the relative orientation between magnetic fields and density structures. The $AM$ for $\phi_{\mathrm{B-N}}$ can be expressed as
\begin{equation} \label{eq:am}
AM = \langle \cos 2\phi_{B-N} \rangle.
\end{equation}
Similar to $Z_x$, the $AM$ is also based on the Rayleigh statistic. $AM$ is in the range of -1 to 1, which can be regarded as a normalized version of $Z_x$. 


\subsection{Observations}\label{sec:hroobs}

The relation between the cloud density structures and magnetic field orientations has been extensively studied observationally. For example, \citet{2013MNRAS.436.3707L} found the orientation of Gould Belt clouds tends to be either parallel or perpendicular to mean orientation of the cloud magnetic field, which they interpreted as strong fields channeling sub-Alfv\'{e}nic turbulence or guiding gravitational contraction. Toward the same sample, clouds elongated closer to the field orientation were found to have (1) higher star formation rates, which was suggested to be due to their smaller magnetic fluxes as well as weaker magnetic support against gravitational collapse \citep{2017NatAs...1E.158L}; (2) shallower mass cumulative function slopes \citep[][]{2020MNRAS.498..850L}, i.e., shallower column density probability distribution functions (N-PDFs), or, in other words, more mass at high densities. In filamentary clouds, there is evidence that the magnetic field is parallel to the low-density striations and is perpendicular to the high-density main filament \citep[e.g.,][]{2011ApJ...741...21C, 2016A&A...590A.110C}, which implies that the main filament accrete gas through the striations along the field lines. Besides the success of those observational studies, the HRO analysis has enabled a way to perform pixel-by-pixel statistics for the local alignment between the column density structure and the magnetic field. Observational studies using the HRO analysis have been focused on studying this alignment at different column densities. 

The first HRO analyses were made with observations from the Planck/HFI at large scales. With a smoothed resolution of 15$\arcmin$, \citet{2016A&A...586A.135P} found that $\xi$ is mostly positive and is anti-correlated with the column density\footnote{A significant amount of the diffuse ISM is in the atomic phase, while molecular clouds are in the molecular phase. Here we use $N_{\mathrm{H_2}}$ to describe the column density in diffuse ISM and molecular clouds for uniformity.} $N_{\mathrm{H_2}}$ over the whole-sky at $N_{\mathrm{H_2}}\lesssim 5\times10^{21}$ cm$^{-2}$. The Planck observations toward 10 nearby ($d<450$ pc) clouds \citep{2016AA...586A.138P} with a smoothed resolution of 10$\arcmin$  ($\sim$0.4-1.3 pc) have revealed a prevailing trend of decreasing $\xi$ with increasing $N_{\mathrm{H_2}}$, with $\xi$ being positive at lower $N_{\mathrm{H_2}}$ and becoming negative at higher $N_{\mathrm{H_2}}$ in most clouds except for those with low column densities (e.g., CrA). The transition of $\xi$ from positive to negative values was found to be at $N_{\mathrm{H_2,tr}} \sim 2.5\times10^{21}$ cm$^{-2}$. 

Subsequent studies have expanded the HRO analysis to compare the large-scale magnetic field observed with Planck/HFI or the BLASTPol with the smaller-scale column density structures revealed by the Herschel Space Observatory. \citet{2016MNRAS.460.1934M} compared the Herschel dust emission structures at a 20$\arcsec$ ($\sim$0.01 pc) resolution and the large-scale magnetic field orientation revealed by Planck polarization maps at a 10$\arcmin$ ($\sim$0.4 pc) resolution and found a trend of decreasing $\xi$ with increasing $N_{\mathrm{H_2}}$ in the nearby cloud L1642, which transits at $N_{\mathrm{H_2,tr}} \sim 8\times10^{20}$ cm$^{-2}$. \citet{2017AA...603A..64S} found the same decreasing $\xi-N_{\mathrm{H_2}}$ trend in the Vela C molecular complex by comparing the large-scale magnetic field orientation revealed by BLASTPol at a smoothed resolution of 3$\arcmin$ ($\sim$0.61 pc) with the column density structures revealed by Herschel at a smoothed resolution of 35.2$\arcsec$ ($\sim$0.12 pc). The transition column density in Vela C is $N_{\mathrm{H_2,tr}} \sim 6\times10^{21}$ cm$^{-2}$. They also found the slope of the $\xi-N_{\mathrm{H_2}}$ relation is sharper in sub-regions where the high-column density tails of N-PDFs are flatter. \citet{2019A&A...629A..96S} compared the Herschel column density maps at a resolution of 36$\arcsec$ ($\sim$0.03-0.08 pc) with the large-scale magnetic field from Planck observations for the ten clouds studied by \citet{2016AA...586A.138P} and found the $\xi$ (or $Z_x$) decreases with increasing $N_{\mathrm{H_2}}$ in most clouds, which is in agreement with the study by \citet{2016AA...586A.138P}. They also found that regions with more negative cloud-averaged $\xi$ (or $Z_x$) tend to have steeper N-PDF tails, but did not find a clear trend between the cloud-averaged $\xi$ (or $Z_x$) and the star formation rate. In addition, \citet{2019ApJ...878..110F} compared the magnetic field orientation revealed by BLASTPol with the integrated line-intensity structures of different molecular lines observed with Mopra in Vela C. They found that the line emission for low-density tracers are statistically more aligned with the magnetic field, while high-density tracers tend to be perpendicular to the magnetic field. The transition occurs at $n_{\mathrm{H_2,tr}} \sim 10^{3}$ cm$^{-3}$. 

At smaller scales, the polarization observations from the SOFIA/HAWC+ or JCMT/POL2 can be used to study the relative orientation between the magnetic field and column density structures within elongated filamentary structures. \citet{2021ApJ...918...39L} compared magnetic field orientations from the HAWC+ observations with the Herschel column density maps in Ophiuchus/L1688 at a smoothed resolution of 36.3$\arcsec$ ($\sim$0.02 pc). They found smaller $\xi$ values at higher $N_{\mathrm{H_2}}$, with $\xi$ being mostly negative except for the first two column density bins. The transition column density in L1688 is $N_{\mathrm{H_2,tr}} \sim 5\times10^{21}$ cm$^{-2}$. They also found that $\xi$ increases from negative to $\sim$0 at higher column densities in the dense core Oph A within L1688, which suggests the magnetic alignment behavior is more complex at higher column densities. \citet{2022MNRAS.510.6085L} compared the inferred magnetic field from HAWC+ with the Herschel column density map in Taurus/B211 and found that the magnetic field is more likely perpendicular to the column density structures in B211. \citet{2022ApJ...926..163K} found negative $\xi$ values in the Serpens Main cloud with JCMT observations at a resolution of 14$\arcsec$ ($\sim$0.03 pc), where they suggested that the first column density bin at $N_{\mathrm{H_2}} \sim 9.3\times10^{21}$ cm$^{-2}$ is approximately the transition column density. In \citet{2022ApJ...926..163K}, the $\xi$ decreases from 0 to negative values from $N_{\mathrm{H_2}} \sim 9.3\times10^{21}$ cm$^{-2}$ to $\sim 4.6\times10^{22}$ cm$^{-2}$, which mainly corresponds to the filamentary structures. Between $N_{\mathrm{H_2}} \sim 4.6\times10^{22}$ cm$^{-2}$ to $\sim 10^{23}$ cm$^{-2}$, $\xi$ increases back to 0, suggesting that the magnetic fields are trending parallel to elongated structures. At $N_{\mathrm{H_2}} > \times10^{23}$ cm$^{-2}$, $\xi$ again decreases with increasing $N_{\mathrm{H_2}}$. This behavior suggests a complex interplay between mass accumulation, hub-filament interaction, and gravitational collapse within filamentary structures. 
 
Interferometric polarization observations are capable of revealing the magnetic field orientation within molecular dense cores. With ALMA observations at a resolution of 0.35$\arcsec$ ($\sim$140 AU), \citet{2017ApJ...842L...9H} found that the magnetic field is slightly more perpendicular than parallel to the column density structure around the class 0 protostellar source Ser-emb 8 in Serpens Main. Using ALMA observations at a resolution of 1.1$\arcsec$ ($\sim$200 AU), \citet{2020ApJ...892..152H} found that the magnetic field tends to be perpendicular to the column density structure in the class 0 protostellar source BHR 71 IRS1. They also found the magnetic field tends to be perpendicular to the column density structure but is parallel to outflow-cavity walls in another class 0 source BHR 71 IRS2, suggesting that the magnetic field is affected by the outflow activity. With ALMA observations at a resolution of 1.2$\arcsec$ ($\sim$0.02 pc) toward the high-mass star-forming region G327.3, \citet{2020ApJ...904..168B} found that the relative orientation between the magnetic field and the dust emission structure changes from perpendicular to a random distribution when it is closer to the dust emission peak (i.e., $\xi$ or $Z_x$ increases with increasing $N_{\mathrm{H_2}}$). This clearly contradicts the general decreasing $\xi-N_{\mathrm{H_2}}$ trend at larger scales found by studies of lower resolution data, and might be related to the rotational properties of potential disks. 

In summary, there is a general trend that the relative orientation between the magnetic field and the column density structure changes from parallel to perpendicular with increasing gas column densities. The transition from parallel to perpendicular alignment occurs at $N_{\mathrm{H_2,tr}} \sim 10^{21}-10^{22}$ cm$^{-2}$ and may be at $n_{\mathrm{H_2,tr}} \sim 10^{3}$ cm$^{-3}$, which are comparable to the transition or critical densities of the $B-n$ and $B-N$ relations measured from DCF estimations \citep{2022ApJ...925...30L} or Zeeman observations \citep{2012ARA&A..50...29C}. Within filaments and molecular dense cores, the relative orientation could return to random alignment in some sub-regions due to the impact of the projection effect (see Section \ref{sec:hrosim}), accreting gas flows, outflows, and/or the disk rotation. More observations are needed to better understand the returning to random alignment at high densities. 


\subsection{Simulations}\label{sec:hrosim}
As indicated in Section \ref{sec:hroobs}, most clouds show a transition from a preferentially parallel alignment between the magnetic field orientation and density structures at low densities to a preferentially perpendicular alignment at higher densities. Numerical HRO studies have been focused on interpreting the contradictory alignment at different densities and understanding the implication of this transition. 

\citet{2013ApJ...774..128S}, \citet{2016AA...586A.138P}, and \citet{2017AA...603A..64S} have analysed 3 self-gravitating models with different initial magnetization levels (plasma $\beta_0$=100, 1, and 0.1). The 3 models are initially super-Alfv\'{e}nic ($\mathcal{M}_{A0}$ = 100, 10, and 3.16). At $\sim2/3$ of the flow crossing time\footnote{\citet{2016AA...586A.138P} did not explicitly indicate the time of the snapshot they studied. We consulted their corresponding author to confirm this information.}, the 3 models become super-Alfv\'{e}nic, trans-Alfv\'{e}nic, and sub-Alfv\'{e}nic on average \citep{2016AA...586A.138P}, respectively. The magnetic field in the 3 models are preferentially parallel to the density structures. Only in the trans- or sub-Alfv\'{e}nic models, the relative orientation between the magnetic field and the density strucutre changes from parallel to perpendicular at high densities \citep{2016AA...586A.138P}. \citet{2017AA...603A..64S} found that the transition occurs when the term $A_{23}$ changes its sign. The $A_{23}$ term is a coefficient in the time-varying equation describing the angle between the magnetic field and density gradient, which is a function of the velocity field, magnetic field, and density distribution. They also found that the local values of $\mathcal{M}_{A}>1$ and $\nabla \boldsymbol{v} < 0$ (i.e., converging gas flow) in high-density regions are associated with perpendicular alignment but are not sufficient for the relative orientation to become perpendicular. 

\citet{2016ApJ...829...84C} analysed 3 self-gravitating models with different initial magnetic levels ($\mathcal{M}_{A0}$ = 2.2, 4.4, and 8.8). The 3 initially super-Alfv\'{e}nic models become trans-to-sub-Alfv\'{e}nic ($\mathcal{M}_{A}$ = 0.78, 0.81, and 0.84) on average due to shock compression when the most evolved core starts to collapse. They suggested that the change of relative orientation from parallel to perpendicular above a transition density happens when the sub-Alfv\'{e}nic gas is gravitationally accelerated to become locally super-Alfv\'{e}nic again in overdense regions. However, \citet{2017ApJ...836...95O} found that the change of relative orientation can still happen even when the high-density self-gravitating region is sub-Alfv\'{e}nic (local $\mathcal{M}_{A}=0.54$) in their simulations. Hence $\mathcal{M}_{A}>1$ may be sufficient but is not necessary for the transition of relative orientation. 

\citet{2020MNRAS.497.4196S} performed a set of self-gravitating colliding flow simulations with $\mathcal{M}_{A0}$ = 0.6-4. At a time snapshot of 19 Myr, the mass-to-flux ratio $\mu$ of the simulations become 0.54-4.3, where the critical $\mu$ is 1. They found that only models with $\mu \lesssim$1 shows a clear transition from parallel to perpendicular alignment for the relative orientation. Note that their models with $\mu \lesssim$1 also have $\mathcal{M}_{A0} \lesssim 1$. Similar to \citet{2017AA...603A..64S}, they suggested that the observed transition can be well explained by the term $A = A_{1}+A_{23}$, where $A_{1}$ is another coefficient describing the evolution of the angle between the magnetic field and density gradient. They found that $\xi<1$ is associated with local $\mathcal{M}_{A}\gg1$ in high density regions, which agrees with \citet{2017AA...603A..64S}. \citet{2020MNRAS.497.4196S} also performed a set of SILCC-Zoom simulations with different physical conditions, but their output are less clear than those of the colliding flow simulations. Therefore no conclusions were drown from these SILCC-Zoom simulations. 

\citet{2020MNRAS.499.4785K} performed a set of initially supersonic (initial Mach number $\mathcal{M}_{s0} = 7.5$) non-gravitating simulations with different initial magnetic levels ($\beta_0$=10, 1, and 0.01; $\mathcal{M}_{A0}$ = 16.7, 5.3, and 0.5), where the turbulence is driven either solenoidally or compressively. They found only models with $\beta_0$=0.01 ($\mathcal{M}_{A0}$ = 0.5) and compressible turbulence clearly show the transition from parallel to perpendicular alignment at higher densities, which may suggest the main driver of the change of relative orientation is the compression of the gas. They also found the transition point for $\xi \sim 0$ does not perfectly correspond to $A = A_{1}+A_{23} \sim 0$, where $A \sim 0$ happens at slightly lower densities than $\xi \sim 0$. Their analysis suggests that $A = A_{1}+A_{23} > 0$ is not a sufficient condition and self-gravity is not a necessary condition for the transition of relative orientation. The change of relative orientation without self-gravity is also seen in the initially trans-to-sub-Alfv\'{e}nic and very supersonic simulations by \citet{2019ApJ...886...17H}.

\citet{2021MNRAS.507.5641G} also found a transition from parallel alignment to perpendicular alignment at higher densities in a 10 pc region of their large-scale simulation. They suggested that the transition happens when the local mass-to-flux ratio exceeds its critical value and the gravitational forces dominates the combination of thermal and magnetic pressures. 

\citet{2021MNRAS.503.5425B} studied the relative orientation between the magnetic field and density structure with simulations with a wide range of initial states (sub-Alfv\'{e}nic and super-Alfv\'{e}nic; with and without gravity). In non-gravitating simulations, only initially sub-Alfv\'{e}nic models with the largest initial sonic number ($\mathcal{M}_{s0} = 7$) show signs of perpendicular alignment between the magnetic field and column density structure in 2D at late stages of evolution. In  simulations with gravity, the initial distribution of $Z_x$ is similar to the simulations without gravity. At later stages, all gravitating sub-Alfv\'{e}nic models show a change of $Z_x$ from positive to negative values at high densities. They concluded that self-gravity may help to create structures perpendicular to the magnetic field. 

\citet{2022ApJ...925..196I} analysed 3 clouds in a galactic-scale simulation with gravity. They found that the gas is trans-Alfv\'{e}nic at $n_{\mathrm{H_2}} < 10^{2}$ cm$^{-3}$. At $n_{\mathrm{H_2}} > 10^{2}$ cm$^{-3}$, the gas becomes super-Alfv\'{e}nic, which they suggested to be due to gravitational collapse. The relative orientation changes from predominantly parallel to much more random or perpendicular at even higher densities. i.e., they also found $\mathcal{M}_{A}>1$ at high density is associated with perpendicular alignment but is not a sufficient condition for the transition of relative orientation, which agrees with \citet{2017AA...603A..64S} and \citet{2020MNRAS.497.4196S}. 

Observations can only trace the projected quantities in the POS. The projection effect can make the 2D $\xi-N_{\mathrm{H_2}}$ relation different from the 3D $\xi-n_{\mathrm{H_2}}$ relation.  Two parallel vectors in 3D are still parallel when projected in 2D, but 2 perpendicular vectors in 3D can have any relative orientations when projected in 2D depending on the viewing angle \citep[e.g.,][]{2016A&A...586A.135P}. \citet{2013ApJ...774..128S} found that the relative orientation preserve well from 3D to 2D in sub-Alfv\'{e}nic environments, except when the magnetic field orientation is close to the LOS. In contrast, the projected relative orientation does not necessarily reflect the relative orientation in 3D in super-Alfv\'{e}nic environments. Other numerical studies found similar results for the projection effect \citep{2020MNRAS.497.4196S, 2021MNRAS.503.5425B, 2021MNRAS.507.5641G}.

The transition density at which the relative orientation changes its sign has also been studied by simulations. There is evidence that models with stronger magnetic field tend to have smaller transition densities in individual numerical studies \citep{2013ApJ...774..128S, 2016ApJ...829...84C}, but this transition may not be solely dependent on the Alfv\'{e}nic Mach number \citep{2022arXiv220311179P}. It should be noted that different numerical studies do not have a consistent way for measuring the local $\mathcal{M}_{A}$. Each study reports the local $\mathcal{M}_{A}$ averaged over arbitrary scales, which makes it challenging to compare their results. 

Table \ref{tab:hro} summarizes the simulations that show the transition of relative orientation from parallel to perpendicular alignment at higher densities. In most simulations \citep{2013ApJ...774..128S, 2016AA...586A.138P, 2017AA...603A..64S, 2016ApJ...829...84C, 2020MNRAS.497.4196S, 2020MNRAS.499.4785K, 2021MNRAS.503.5425B, 2022ApJ...925..196I}, the transition of relative orientation only occurs when the large-scale environment is trans- or sub-Alfv\'{e}nic.  For non-self-gravitating simulations, the transition only happens when the initial environment is sub-Alfv\'{e}nic and supersonic \citep{2020MNRAS.499.4785K, 2021MNRAS.503.5425B}. Alternatively, \citet{2020MNRAS.497.4196S} proposed that the transition only occurs in an magnetically sub-critical large-scale environment. It is still unclear what triggers the relative orientation to change from parallel to perpendicular. Plausible reasons include local $\mathcal{M}_{A}>1$\footnote{It is unclear whether these numerical studies has excluded the non-turbulent motions (e.g., infall, outflow, and/or rotation.) in the calculation of $\mathcal{M}_{A}$, so the $\mathcal{M}_{A}>1$ at high densities found in these simulations may be arguable.}, $A_{1}+A_{23}>0$, and/or $\nabla \boldsymbol{v} < 0$ in high-density regions, which are associated with the perpendicular alignment but may not be sufficient conditions for the transition. Alternatively, local $\mu>1$ in high-density regions may also be responsible for the transition, where a dominant gravity can help to create perpendicular alignment between the magnetic field and density structure. 

The analytical explanation for the parallel alignment between the magnetic field and density structure at low densities and the perpendicular alignment at high densities is still unclear. At low densities, the gravity is not dominant, so the alignment is mainly due to the interplay between the magnetic field and turbulence. \citet{2013ApJ...774..128S}, \citet{2016A&A...586A.135P}, \citet{2016AA...586A.138P}, and \citet{2017AA...603A..64S} proposed that the initially super-Alfv\'{e}nic compressive turbulent flows can stretch the magnetic field and density structures in the same direction due to magnetic freezing. However, the parallel alignment is also seen in initially sub-Alfv\'{e}nic simulations (see Table \ref{tab:hro}). So an initially super-Alfv\'{e}nic enviroment might not be a necessary condition for the parallel alignment. Alternatively, \citet{2016ApJ...829...84C} and \citet{2019ApJ...878..110F} suggested that the anisotropic turbulent eddies in sub-Alfv\'{e}nic gas naturally create elongated density structures
parallel to the magnetic field. The anisotropic nature of sub-Alfv\'{e}nic turbulence can explain the majority of the low-density gas with local $\mathcal{M}_{A}<1$ and $\xi>0$, but cannot explain the faction of gas with local $\mathcal{M}_{A}>1$ and $\xi>0$ in some simulations \citep{2017AA...603A..64S, 2020MNRAS.497.4196S, 2022ApJ...925..196I}. At high densities, the situation is more complicated as the gravity steps in. \citet{2013ApJ...774..128S} and \citet{2016AA...586A.138P} suggested that the gas compression along field lines in sub-Alfv\'{e}nic turbulence can create density structures perpendicular to the magnetic field even without gravity. Alternatively, a magnetized gravitational collapse in the presence of a strong magnetic field would naturally cause an elongated density structure that is perpendicular to the magnetic field \citep{1976ApJ...206..753M, 1976ApJ...207..141M}. However, both explanations are inconsistent with the results of some simulations where $\mathcal{M}_{A}>1$ is associated with the perpendicular alignment. Another alternative explanation is that the perpendicular alignment at high densities may be determined by the strong large-scale magnetic field at low densities rather than the small-scale magnetic field at high densities \citep{2016A&A...586A.135P}. 

\begin{table}[tbh]
\tiny
\caption{Simulations that show transitions of relative orientation from parallel to perpendicular alignment at higher densities.  \label{tab:hro}}
\begin{tabular}{ccccccccc}
\hline \noalign {\smallskip}
Size (pc)  & Gravity & $\mu_0$ & $\mathcal{M}_{A0}$ & $\mu$ & $\mathcal{M}_{A}$ & $\xi>0$$^d$ & $\xi<0$$^d$ &  Ref.$^e$ \\
\hline \noalign {\smallskip}
4 & Yes & 4.52-14.3 & 3.16-10 & ... & $\lesssim$1 & $A_{23}<0$ & $A_{23}>0$, $\mathcal{M}_{A}>1$, $\nabla \boldsymbol{v} < 0$ & 1,2,3\\ 
1 & Yes & ... & 2.2-8.8 & ... & 0.78-0.84 & $\mathcal{M}_{A}<1$ & $\mathcal{M}_{A}>1$ & 4 \\ 
32$^a$ & Yes & ... & 0.6-0.8 & 0.54-0.72 & ... & $A_{1}+A_{23}<0$ & $A_{1}+A_{23}>0$,$\mathcal{M}_{A}>1$ & 5 \\ 
10 & No & ... & 0.5 & ... & ... & ... & $A_{1}+A_{23}>0$ & 6 \\ 
10$^b$ & Yes & ... & ... & ... & ... & $\mu<1$ & $\mu>1$ & 7 \\ 
10 & No & ... & 0.7 & ... & ... & ... & ... & 8 \\ 
10 & Yes & ... & 0.6 & ... & ... & ... & ... & 8 \\ 
100$^c$ & Yes & ... & 0.6 & ... & ... & ... & $\mathcal{M}_{A}>1$ & 9 \\ 

\hline \noalign {\smallskip}
\end{tabular}

\normalsize{Notes}\\
\normalsize{$^{a}$ The colliding flow simulations in \citet{2020MNRAS.497.4196S} have a box volume of 128 $\times$ 32 $\times$ 32 pc$^3$. They only selected 32$^3$ pc$^3$ regions for study.}\\
\normalsize{$^{b}$ The simulations in \citet{2021MNRAS.507.5641G} have a box volume of 500$^3$ pc$^3$. They only selected 10$^3$ pc$^3$ regions for study.}\\
\normalsize{$^{c}$ The simulations in \citet{2022ApJ...925..196I} have a box volume of 1 $\times$ 1 $\times$ 40 kpc$^3$. They only selected 100$^3$ pc$^3$ regions for study.}\\
\normalsize{$^{d}$ The parameters listed in the two columns correspond to local values in sub-regions of $\xi>0$ or $\xi<0$ at different densities.}\\
\normalsize{$^{e}$ References: (1) \citet{2013ApJ...774..128S};  (2) \citet{2016AA...586A.138P}; (3) \citet{2017AA...603A..64S}; (4) \citet{2016ApJ...829...84C}; (5) \citet{2020MNRAS.497.4196S}; (6) \citet{2020MNRAS.499.4785K}; (7) \citet{2021MNRAS.507.5641G}; (8) \citet{2021MNRAS.503.5425B}; (9) \citet{2022ApJ...925..196I}.}\\
\end{table}

\section{The KTH method}\label{sec:KTH}

\subsection{Basics}
\citet{2012ApJ...747...79K} proposed the KTH method to determine the magnetic field strength with the relative orientations between the magnetic field, emission intensity gradient, and local gravity. This method is based on ideal MHD force equations, where the intensity gradient is assumed to trace the resulting direction of motions in the MHD equation \citep[i.e., the inertial term, ][]{2013ApJ...775...77K}. This method leads to maps of position-dependent magnetic field strengths and magnetic-to-gravitational force ratios. 

Under the assumptions of negligible viscosity, infinite conductivity (ideal MHD), isotropic magnetic field pressure, small turbulent-to-ordered field strength ratio, smoothly and slowly varying field strength, stationarity, and that the intensity gradient indicates the resulting direction of motions, \citet{2012ApJ...747...79K} considered the force balance among pressure, gravity, and the curvature term of magnetic field to derive the field strength
\begin{equation}\label{eq:BKTH}
    B = \sqrt{\frac{\sin \psi}{\sin \alpha}(\nabla P + \rho \nabla \phi_G) 4\pi R_B},
\end{equation}
where $\psi$ is the relative orientation between the gravity and the intensity gradient, $\alpha$ is the relative orientation between the magnetic field and the intensity structure\footnote{The angle $\alpha$ is equivalent to the $\phi_{\mathrm{B-N}}$ in Section \ref{sec:hro} if the intensity structure perfectly traces the density structure.}, $P$ is the hydrostatic dust pressure, $\phi_G$ is the local gravitational potential, and $R_B$ is the local magnetic field line radius. \citet{2012ApJ...747...79K} then introduced a parameter $\Sigma_B$ to measure the local field significance. The field significance parameter $\Sigma_B$ is given by the reformulation of Equation \ref{eq:BKTH}
\begin{equation}\label{eq:sigmaB}
    \Sigma_B = \frac{\sin \psi}{\sin \alpha} = \frac{F_B}{\vert F_G + F_P \vert},
\end{equation}
which quantifies the relative importance of the magnetic force $F_B = B^2/(4\pi R_B)$ compared to the combination of the gravitational and pressure forces $F_G + F_P = \nabla P + \rho \nabla \phi_G$. \citet{2012ApJ...747...79K} further demonstrated analytically that Equations \ref{eq:BKTH} and \ref{eq:sigmaB} do not suffer too much from the geometry and projection effects. If local changes in temperatures and densities are small compared to gravity, the pressure terms ($\nabla P$ and $F_P$) can be omitted from Equations \ref{eq:BKTH} and \ref{eq:sigmaB}. Thus, $\Sigma_B$ can be used to quantitatively indicate whether the magnetic field in a region is strong enough to prevent gravitational collapse ($\Sigma_B>1$) or not  ($\Sigma_B<1$). Later, \citet{2012ApJ...747...80K} suggested that the global mass-to-flux ratio normalized to the critical value within a specific region can be estimated with
\begin{equation}\label{eq:lambdaKTH}
    \lambda_{\mathrm{KTH}} = \langle \Sigma_B^{-1/2} \rangle \pi^{-1/2},
\end{equation}
where they adopted the magnetic critical mass from \citet{1999ASIC..540..193S} in the derivation of $\lambda_{\mathrm{KTH}}$. 

Due the a series of assumptions, the KTH method is subjected to many uncertainties \citep{2012ApJ...747...79K}. In contrast to the DCF method that requires the information on the turbulent field, the KTH method regards the turbulent field as insignificant and only requires the ordered field structure. The contribution from turbulent field may be removed by averaging several neighboring pixels to derive an averaged local ordered field curvature (e.g., applying the Pil15 or Pat17 technique mentioned in Section \ref{sec:unorder}), but the KTH method may still fail when the magnetic field is dominated by the turbulent component. The KTH method also is not applicable in the extremely strong field case where the matter can move only along the field lines (i.e., $\psi \sim 0$, $\alpha \sim \pi/2$, and $R_B \sim \infty$). In some regions with strong rotation, the effect of rotation can lead to a change in the resulting direction of motion, which may be mitigated by adding a centrifugal term in the MHD equation. If the temperature variation is irregular and significant over a map, the intensity gradient should be replaced by the density gradient in the KTH analysis. The biggest uncertainty of the KTH method may come from the basic assumption that the intensity (or density) gradient traces the resulting motion of the MHD force equation. \citet{2013ApJ...775...77K} has investigated this assumption analytically and suggested that its validity relies on the difference between the velocity and density gradient directions. Although the numerical studies of the HRO method can be regarded as a partial test of the KTH method, the KTH method itself has not been fully compared with simulations yet. Further numerical investigations will be of significance to understand the uncertainty of this method in different physical conditions.



\subsection{Observations}

Most observational studies using the KTH method only partially applied this method and studied the distribution of the intermediate parameters. Here we summarise the observational studies of each parameter. 

\subsubsection{Magnetic field versus intensity gradient: $\delta$} \label{sec:KTHdelta}
The angle $\delta = 90\degr - \alpha$ is the relative orientation between the magnetic field and the intensity (or density) gradient. Basically, studying the distribution of $\delta$ is similar to studying the distribution of $\phi_{\mathrm{B-N}}$ introduced by the HRO technique. 

Similar to $\phi_{\mathrm{B-N}}$, the interpretation of the value and distribution of $\delta$ is still not yet well established. \citet{2013ApJ...775...77K} and \citet{2014ApJ...797...99K} suggested that a bimodal distributions of $\delta$ can be interpreted as a sign of collapse, where the angle $\delta$ measures how efficiently the magnetic field inhibits a collapse. \citet{2013ApJ...775...77K} and \citet{2014ApJ...797...99K} further proposed that the angle $\delta$ can used as a tracer of the evolution stage of star-forming regions. They suggested that $\delta$ is spatially randomly distributed due to the lack of a gravitational center in the early phase (Type I). In a later stage, elongated dust structures appear in star-forming regions. The magnetic field is parallel to the major axis of elongated dust structures that are created by large-scale flows and/or turbulence (Type IIB) or is perpendicular to the major axis of elongated dust structures when the gravity has just started to shape the field (Type IIA). Types IIB and IIA may further evolve into one and the same system (Type III) where a dominant gravity drags the magnetic field to form a radial, pinched, or hourglass shape. \citet{2014ApJ...797...99K} analysed a sample of 50 star-forming regions with CSO and SMA observations and found that the mean $\vert \delta \vert$ values for Types IIB, IIA, and III are 51$\degr$ (CSO and SMA), 30$\degr$ (CSO) to 34$\degr$ (SMA), and 30$\degr$ (SMA only), respectively. The categorization of different $\delta$ types is an empirical characterization. The correspondence between the $\delta$ types and different evolution stages of low-mass and high-mass star formation at different scales is still unclear and warrants further investigations. 

Studies of individual star-forming regions have revealed different values and distributions of $\delta$. \citet{2013ApJ...763..135T} found that the magnetic field is less correlated with the dust intensity gradient at larger scales revealed by JCMT or CSO, while the two angles are more correlated at smaller scales observed by SMA. This trend is consistent with the results from the observational HRO studies (Section \ref{sec:hroobs}). With SMA polarization observations toward the massive dense cores in the DR21 filament, \citet{2017ApJ...838..121C} found that the magnetic field and the intensity gradient are misaligned in rotation-like cores and are aligned in non-rotation-like cores, which suggests the magnetic field could be distorted by the rotation. \citet{2013ApJ...772...69G} found random distributions of $\delta$ with an average value of 40$\degr$ in the massive dense core DR21(OH) with SMA observations. With ALMA polarization observations toward the massive clumps in the IRDC G28.34, \cite{2020ApJ...895..142L} found random distributions of $\delta$ and average $\delta$ values of 40$\degr$ and 46$\degr$ in the dense cores in two massive clumps MM1 and MM4, respectively. In another IRDC G14.225, \citet{2020A&A...644A..52A} found that the $\delta$ value in the Hub-N region is mostly small with their CSO observations. More observational studies of $\delta$ are still required to better understand its general distributions at different scales and evolution stages of star formation.


\subsubsection{Magnetic field versus local gravity: $\omega$}
The angle $\omega$ measures the relative orientation between the magnetic field and the local gravity. The angle $\omega$ may be used as a measure for how effectively gravity can shape the magnetic field structure \citep{2012ApJ...747...79K}, but further explanations of the distribution of $\omega$ are still yet to be established. 

There are few observational studies of $\omega$. \citet{2012ApJ...747...79K} found an average $\omega$ of 13$\degr$ in W51 e2, suggesting the magnetic field lines are mainly dragged by gravity in this core. With ALMA polarization observations, \citet{2018ApJ...855...39K} found average $\sin \omega$ values of 0.4, 0.41, and 0.47 (i.e., $\omega \sim$ 23$\degr$, 23$\degr$, and 27$\degr$) in three massive cores W51 e2, W51 e8, and W51N, respectively. The distributions of $\omega$ in three cores vary, but all show some magnetic channels with $\sin \omega \sim 0$. \citet{2018ApJ...855...39K} suggested that the gas collapse can proceed in free-fall in these magnetic channels without magnetic resistance, which may have some interesting implication for the star formation rate. \cite{2020ApJ...895..142L} found average $\omega$ values of 34$\degr$ and 36$\degr$ in the dense cores in massive clumps G28-MM1 and G28-MM4, respectively, which are lower than the average $\delta$ values in the same region. This shows that the magnetic field is more aligned with the gravity direction than the intensity (or density) gradient, which suggests $\omega$ may be better than $\delta$ (or $\phi_{\mathrm{B-N}}$ of the HRO technique) for the study of the correlation between the magnetic field and gravity in high density regions. An interesting future study point of $\omega$ may be studying its variation at different densities in a similar way to the studies of the angle $\phi_{\mathrm{B-N}}$. 

\subsubsection{Intensity gradient versus local gravity: $\psi$}
The angle $\psi$ measures the relative orientation between the intensity (or density) gradient and the local gravity. The angle $\psi$ may indicate how effective gravity can shape the density structure, but further analytical implications of $\psi$ are still yet to be investigated.

There are few observational studies of the angle $\psi$. \citet{2012ApJ...747...79K} found an average $\psi$ of 20$\degr$ in W51 e2. \cite{2020ApJ...895..142L} found average $\psi$ values of 30$\degr$, 22$\degr$, and 28$\degr$ in the dense cores in massive clumps G28-MM1, G28-MM4, and G28-MM9, respectively. These observations suggest that the gravity direction is closely aligned with the intensity gradient in high density molecular cores. 

\subsubsection{Magnetic force versus gravitational force: $\Sigma_B$}
The field significance parameter $\Sigma_B$ measures the ratio between the magnetic force and the gravitational force if the gas pressure is negligible. The implication of $\Sigma_B$ is clear, but the accuracy of the estimated force ratio is not tested by simulations yet. 

Based on a sample of 50 sources, \citet{2014ApJ...797...99K} found that the different types of star-forming regions categorized by the angle $\delta$ (see Section \ref{sec:KTHdelta}) show clear segregation in $\Sigma_B$ values. Type IIB sources where the magnetic field is aligned with the clump/core major axis have an average $\Sigma_B$ of 1.29 (CSO) to 1.49 (SMA), which suggests that type IIB sources are supported by the magnetic field and do not collapse on average. Type IIA sources where the magnetic field is perpendicular with the clump/core major axis have an average $\Sigma_B$ of 0.69 (CSO) to 0.74 (SMA), which suggests that type IIA sources are collapsing on average. Type III sources at a later stage have a smaller average $\Sigma_B$ value of 0.59 (SMA only), suggesting an even more dynamical collapsing. 

Individual studies have mostly found $\Sigma_B\lesssim1$ in star-forming dense clumps/cores \citep[e.g., W51 e2, W51A, W51N, DR21(OH), clumps in G34.43, cores in G28.34, and clumps in G14.225, ][]{2012ApJ...747...79K, 2012ApJ...747...80K, 2013ApJ...763..135T, 2013ApJ...772...69G, 2019ApJ...878...10T, 2020ApJ...895..142L, 2020A&A...644A..52A}. Specifically, \citet{2013ApJ...763..135T} found that the lower density structures in W51N revealed by the CSO and JCMT have $\Sigma_B\sim0.71-1.17$, while the higher density structures revealed by SMA have $\Sigma_B\sim0.5$. \citet{2012ApJ...747...79K} and \citet{2012ApJ...747...80K} found smaller $\Sigma_B$ values in higher column density regions in W51 e2 and W51A. These findings indicate that the gravity plays an more dominant role in higher density regions, which agrees with the observational DCF studies (see Section \ref{sec:dcflambda}). On the other hand, \citet{2020ApJ...895..142L} found that the $\Sigma_B$ values are higher in more evolved dense cores in the IRDC G28.34, which might suggest a more dynamical star formation at earlier evolution stages. 

\subsubsection{Magnetic field strength}
The KTH method can be used to estimate the magnetic field strength with Equation \ref{eq:BKTH}. The accuracy of the estimated field strength is still unclear as it has not been tested by simulations yet. 

By far, there are only two observational studies that has applied the KTH method to estimate the field strength. \citet{2012ApJ...747...79K} found an average field strength of 7.7 mG in W51 e2. \citet{2013ApJ...769L..15S} estimated a field strength of 3.4 mG with the KTH method in L1157-mm, which is $\sim$2.5 times higher than the DCF estimation of 1.4 mG in the same work. 

\subsubsection{Mass-to-flux-ratio to critical value}
The mass-to-flux-ratio to critical value $\lambda_{\mathrm{KTH}}$ can be estimated from the field significance parameter $\Sigma_B$ through Equation \ref{eq:lambdaKTH}. Similarly, the accuracy of $\lambda_{\mathrm{KTH}}$ is yet to be investigated by simulations. 

\citet{2012ApJ...747...80K} found larger $\lambda_{\mathrm{KTH}}$ values in higher density regions in W51 e2, which agrees with the trend from DCF estimations (see Section \ref{sec:dcflambda}). Other two observational studies in DR21(OH) and G14.225N have found that the $\lambda_{\mathrm{KTH}}$ value estimated with the KTH method approximately agrees with the value of DCF estimations \citep{2013ApJ...772...69G, 2020A&A...644A..52A}. 

\section{Summary}\label{sec:sum}
The recent improvement of instrumental sensitivity and development of new techniques (e.g., the VGT) have led to an increasing number of observations that reveal the POS component of magnetic field orientations in star-forming molecular clouds. In this review, we discuss the developments and limitations of the DCF and KTH methods that quantify the dynamic importance of magnetic fields in star-forming molecular clouds based on the field orientations and the HRO analysis that characterize the statistical relation between field orientations and density structures.  We also summarize the observational studies using these methods and discuss their implications on star formation. 

The original DCF method is based on several assumptions: the total magnetic field is composed of a mean field component and a turbulent field component; the energy equipartition; isotropic turbulence; and the turbulent-to-mean or -total field strength ratio is traced by angular dispersions. The ordered field component is considered instead of the mean field component (e.g., the ADF method) if there are curved ordered magnetic field structures. We suggest that the ordered field and turbulent field of a particular region are local properties and are dependent on the scale range (i.e., the beam resolution to the maximum recoverable scale of interferometric observations or the region size) of the region of interest. There is still a debate on whether there is an equipartition between the turbulent magnetic field and the turbulent kinetic field or between the coupling-term magnetic field and the turbulent kinetic field in the sub-Alfv\'{e}nic regime,  while both equipartitions are not satisfied for super-Alfv\'{e}nic turbulence. The energy non-equipartition can be the biggest uncertainty in the DCF method, which should be further investigated with simulations and observations. The uncertainty from anisotropic turbulence may be insignificant for the DCF estimations in self-gravitating regions. The turbulent-to-underlying or -total field strength ratio can be expressed as different forms of angular dispersions, but each has its limitations. The ADF method correctly accounts for the beam-smoothing effect, interferometric filtering effect, and the ordered field structure, but its applicability for quantifying the turbulent field and LOS signal integration may need further numerical investigations. The correction factor for the most widely used formula $B^{\mathrm{u}}_{\mathrm{pos}} \sim \sqrt{\mu_0 \rho }\frac{\delta v_{\mathrm{los}}}{\delta\phi_{\mathrm{obs}}}$ decreases with increasing density. The DMA presents an important improvement for the DCF method by analytically accounting for the mean field inclination angle and the turbulence anisotropy in the non-self-gravitating regime. A further extension of the DMA in the self-gravitating regime would be of significance. 

A compilation of previous DCF estimations suggests a scenario that magnetically sub-critical low-density clouds gradually form super-critical high-density substructures. The critical column density is around $3.4 \times 10^{21}$ cm$^{-2}$ on average, which needs to be better constrained and may differ in different clouds. The gravity may be more dominant in high-mass star formation than low-mass star formation. The average state of dense substructures within molecular clouds is approximately trans-Alfv\'{e}nic if the energy equipartition assumption is satisfied, or super-Alfv\'{e}nic if the energy equipartition assumption is unsatisfied for some of the sources. 

Observational HRO studies mainly focus on the alignment between the magnetic field and density structure. Low-resolution HRO studies have found a general trend of transition from a preferentially parallel alignment at low column densities to a perpendicular alignment at higher column densities. This observational trend agrees with trans-to-sub-Alfv\'{e}nic simulations, which indicates that the star-forming molecular clouds are trans-to-sub-Alfv\'{e}nic. This trans-to-sub-Alfv\'{e}nic state is consistent with the results derived from other techniques \citep[e.g., the VGT, ][]{2019NatAs...3..776H}. The analytical explanation for the transition from parallel to perpendicular alignment is still unclear, but may be related to changes of the local Alfv\'{e}nic Mach number, $A_{1}+A_{23}$ term, mass-to-flux-ratio, and/or $\nabla \boldsymbol{v}$. The transition occurs at $10^{21}- 10^{22}$ cm$^{-2}$, which agrees with the critical column density derived from DCF estimations. But it is unclear whether the two transition column densities are related. High-resolution HRO studies have revealed a possible transition from perpendicular alignment back to random alignment at high column density sub-regions. The reason for this reverse transition is also unclear, but may be related to the impact of accretion gas flows, outflows, disk rotation, and/or the projection effect. 

The advantage of the KTH method compared to the DCF method is that it does not require the information on the velocity dispersion. However, the uncertainty of the KTH method is still unknown since it has not been fully tested by simulations. Results from observational KTH studies on the relative alignment between the magnetic field and intensity (density) gradient within dense clumps/cores approximately agree with those of the observational HRO studies. The value and density-varying trend of the mass-to-flux-ratio and the magnetic field strength derived from the KTH method approximately agree with those derived from the DCF estimations. 

\section*{Conflict of Interest Statement}

The authors declare that the research was conducted in the absence of any commercial or financial relationships that could be construed as a potential conflict of interest.

\section*{Author Contributions}
J.L., Q.Z. and K.Q. contributed to the outline of the review. J.L. led the writing of the manuscript. Q.Z. and K.Q. read, commented on, and edited the manuscript. 


\section*{Funding}
J.L. acknowledges the support from the EAO Fellowship Program under the umbrella of the East Asia Core Observatories Association. K.Q. is supported by National Key R\&D Program of China grant No. 2017YFA0402600. K.Q. acknowledges the support from National Natural Science Foundation of
China (NSFC) through grant Nos. U1731237, 11590781, and 11629302.

\section*{Acknowledgments}
We thank the referees for constructive comments
that improved the clarity of this paper. J.L. thanks Prof. Martin Houde for insightful discussions about the correlation between turbulent magnetic and kinetic fields. J.L. thanks Dr. Heshou Zhang and Dr. Suoqing Ji for helpful discussions on the general concept of MHD turbulence. 



\bibliographystyle{frontiersinSCNS_ENG_HUMS} 
\bibliography{astro}




\end{document}